\documentclass[12pt]{article}
\def\Slash#1{#1\kern-0.55em\raise.05ex\hbox{/}}
\usepackage{amsmath,amssymb,amsfonts,amscd,bbm,cancel,graphicx,hyperref,longtable}
\newcommand{\eq}{Eq.}

\newcommand{\fig}{Fig.}

\newcommand{\Ref}{Ref.}

\newcommand{\Tab}{Table}

\newcommand{\ba}{\begin{array}}
\newcommand{\ea}{\end{array}}
\newcommand{\be}{\begin{equation}}
\newcommand{\ee}{\end{equation}}
\newcommand{\equ}[1]{\eq~(\ref{equ:#1})}

\newcommand{\figu}[1]{\fig~\ref{fig:#1}}

\begin{document}


\thispagestyle{empty}
\renewcommand{\thefootnote}{\fnsymbol{footnote}}
\setcounter{footnote}{1}

\vspace*{-0.5cm}

\centerline{\Large\bf Group Space Scan of Flavor Symmetries for}
\vspace*{2mm}
\centerline{\Large\bf Nearly Tribimaximal Lepton Mixing}

\vspace*{18mm}

\centerline{\large\bf
Florian Plentinger\footnote{E-mail:
  \texttt{fplentinger@physik.uni-wuerzburg.de}},
Gerhart Seidl\footnote{E-mail: \texttt{seidl@physik.uni-wuerzburg.de}},
and Walter Winter\footnote{E-mail: \texttt{winter@physik.uni-wuerzburg.de}}
}

\vspace*{5mm}
\begin{center}
{\em Institut f\"ur Theoretische Physik und Astrophysik}\\
{\em Universit\"at W\"urzburg, D-97074 W\"urzburg, Germany} 
\end{center}

\vspace*{20mm}

\centerline{\bf Abstract}
\vspace*{2mm}
We present a systematic group
space scan of discrete Abelian flavor symmetries for lepton mass
models that produce nearly tribimaximal lepton
mixing. In our models, small neutrino masses are generated by
the type-I seesaw mechanism. The lepton mass matrices
emerge from higher-dimension operators via the Froggatt-Nielsen mechanism and are
predicted as powers of a single expansion parameter $\epsilon$ that is
of the order of the Cabibbo angle $\theta_\text{C}\simeq 0.2$. We focus on solutions that can give close to tribimaximal
lepton mixing with a very small reactor angle
$\theta_{13}\approx 0$ and find several thousand explicit such models
that provide an excellent fit to current neutrino data. The models are rather general in the sense
that large leptonic mixings can come from the charged leptons
and/or neutrinos. Moreover, in the
neutrino sector, both left- and right-handed neutrinos can mix
maximally. We also find a new relation
$\theta_{13}\lesssim\mathcal{O}(\epsilon^3)$ for the reactor angle and
a new sum rule $\theta_{23}\approx\frac{\pi}{4}+\epsilon/\sqrt{2}$ for the
atmospheric angle, allowing the models to be tested in future neutrino
oscillation experiments.

\renewcommand{\thefootnote}{\arabic{footnote}}
\setcounter{footnote}{0}

\newpage


\section{Introduction}
\label{sec:introduction}

During the past decade, solar~\cite{Fukuda:2002pe},
atmospheric~\cite{Fukuda:1998mi}, reactor~\cite{Araki:2004mb}, and
accelerator~\cite{Aliu:2004sq} neutrino oscillation experiments, have
very well established that neutrinos are massive. Since neutrinos are
massless in the standard model (SM), massive neutrinos signal physics beyond the SM. In fact, the smallness of
neutrino masses $\sim 10^{-2}\dots 10^{-1}\:\text{eV}$ can be
naturally connected with grand unified theories (GUTs) \cite{SU5,Pati:1974yy} via the seesaw mechanism
\cite{typeIseesaw,typeIIseesaw}, in which the absolute neutrino mass
scale becomes suppressed by an energy scale close to the GUT scale $M_\text{GUT}\approx 2\times
10^{16}\text{GeV}$ \cite{GUTscale}.

Current neutrino oscillation data (for a recent global fit see
Ref.~\cite{Schwetz:2006dh}) tells us that the Pontecorvo-Maki-Nakagawa-Sakata (PMNS) lepton mixing matrix
$U_\text{PMNS}$ \cite{PMNS} can be approximated by the Harrison-Perkins-Scott (HPS)
tribimaximal mixing matrix $U_\text{HPS}$
\cite{Harrison:1999cf} as
\begin{equation}\label{eqU:HPS}
U_\text{PMNS}\approx U_\text{HPS}=
\left(
\begin{matrix}
\sqrt{\frac{2}{3}} & \frac{1}{\sqrt{3}} & 0\\
-\frac{1}{\sqrt{6}} & \frac{1}{\sqrt{3}} & -\frac{1}{\sqrt{2}}\\
-\frac{1}{\sqrt{6}} & \frac{1}{\sqrt{3}} & \frac{1}{\sqrt{2}}
\end{matrix}
\right).
\end{equation}
In $U_\text{HPS}$, the solar angle $\theta_{12}$ and the atmospheric
angle $\theta_{23}$ are given by $\theta_{12}=\text{arctan}(1/\sqrt{2})$ and
$\theta_{23}=\pi/4$, whereas
the reactor angle $\theta_{13}$ vanishes, i.e., $\theta_{13}=0$. The
actually observed leptonic mixing angles in $U_\text{PMNS}$ may
then be expressed in terms of deviations from tribimaximal mixing
\cite{Plentinger:2005kx,Majumdar:2006px} as ``nearly'' or ``near'' tribimaximal lepton mixing \cite{Xing:2002sw}.

Many models have been proposed in the literature to reproduce
tribimaximal leptonic mixing using non-Abelian flavor symmetries (for
early models based on $A_4$ and examples using the double covering
group of $A_4$, see Refs.~\cite{A4} and \cite{2A4}). These models,
however, have generally difficulties (for a discussion see Ref.~\cite{Altarelli:2007cd}) to predict the observed fermion
mass hierarchies and the Cabibbo-Kobayashi-Maskawa (CKM) quark mixing
matrix $V_\text{CKM}$ \cite{CKM} (see Ref.~\cite{discretequarks} for models
including quarks and Ref.~\cite{discreteGUTs} for unified models). An
interesting connection between the quark and the lepton sector, on the
other hand, is implied by the idea of quark-lepton complementarity
(QLC) \cite{qlc}, which is the phenomenological observation that the
measured solar mixing
angle very accurately satisfies the relation
$\theta_{12}+\theta_\text{C}\approx\pi/4$, where
$\theta_\text{C}\simeq 0.2$ is the Cabibbo angle. QLC has been studied
from many different points of view: as a correction to bimaximal
mixing~\cite{qlcbimax}, together with sum rules \cite{qlcsumrules},
with stress on phenomenological aspects \cite{qlcpheno}, in
conjunction with
parameterizations of $U_\text{PMNS}$ in terms of
$\theta_\text{C}$~\cite{qlcCabibbo}, with respect to statistical
arguments \cite{Chauhan:2006im}, by including renormalization group
effects \cite{qlcRG}, and in model building
realizations~\cite{qlcmodels}.

In Refs.~\cite{Plentinger:2006nb,Plentinger:2007px}, we have suggested
a generalization of QLC to ``extended QLC'' (EQLC), where the mixing
angles of the left- and right-handed leptons can assume any of the values
$\frac{\pi}{4},\epsilon,\epsilon^2,0$. Here, $\epsilon$ is of the
order of the Cabibbo angle $\epsilon\simeq 0.2$. By expressing also the lepton mass
ratios as powers of $\epsilon$, we have derived in
Ref.~\cite{Plentinger:2007px} for the CP-conserving case (a discussion
of nonzero phases can be found in Ref.~\cite{Winter:2007yi}) in total 1981
qualitatively distinct mass matrix textures for the
charged leptons and neutrinos that lead to nearly
tribimaximal neutrino mixing with a small reactor angle
$\theta_{13}\approx 0$. For these textures, the neutrino masses
become small due to
the canonical type-I seesaw mechanism (for a related approach see
Ref.~\cite{Branco:2007nn}). The matrix elements of these textures are
in the flavor basis all expressed by powers of $\epsilon$, which
serves as a single small expansion parameter of the matrices. This
suggests a model building interpretation of the textures in terms of
flavor symmetries, e.g. via the Froggatt-Nielsen mechanism \cite{Froggatt:1978nt}.

In this paper, we describe the systematic construction of several thousand explicit lepton mass models, in which nearly tribimaximal
lepton mixing and the mass hierarchies of charged
leptons and neutrinos emerge from products of discrete Abelian flavor
symmetries. Our motivation is that models with Abelian flavor symmetries generally have the merit
that they need only a very simple scalar sector to achieve the
necessary flavor symmetry breaking. We perform a group space scan of products of discrete $Z_n$ flavor
symmetries using the results of EQLC from
Ref.~\cite{Plentinger:2007px} and restrict to the case of real lepton mass
matrices, i.e., we consider the CP-conserving case. In order to generate small neutrino masses, we assume only the canonical type-I seesaw mechanism. The hierarchical pattern of the lepton mass matrices
results from higher-dimension operators that are produced by the
Froggatt-Nielsen mechanism. Moreover, we are interested only in flavor
symmetries that yield nearly tribimaximal lepton mixing with a very small reactor angle
$\theta_{13}\approx 0$.

The paper is organized as follows: In Sec.~\ref{sec:masses+mixings},
we introduce the notation for the lepton masses and mixings. Next, in
Sec.~\ref{sec:flavorsymmetries}, we specify our discrete flavor
symmetries and describe the generation of lepton mass terms via the
Froggatt-Nielsen mechanism. Then, we outline in
Sec.~\ref{sec:selection} our approach to the group space scan of flavor symmetries. Our general results, a list of explicit example
models, a new relation for the reactor angle, and our sum rules for
the PMNS angles including a new sum rule for the atmospheric angle, are shown in Sec.~\ref{sec:scan}. Finally, we present in Sec.~\ref{sec:summary} our summary and conclusions.

\section{Lepton Masses and Mixings}\label{sec:masses+mixings}
We assume the SM with gauge group $G_\text{SM}=SU(3)_c\times
SU(2)_L\times U(1)_Y$ plus three right-handed neutrinos that
generate small neutrino masses via the type-I seesaw mechanism \cite{typeIseesaw}. The lepton Yukawa couplings and mass terms are
\begin{equation}\label{equ:Yukawas+Masses}
 \mathcal{L}_\text{Y}=
-(Y_\ell)_{ij}H^\ast\ell_ie^c_j-(Y_D)_{ij}\text{i}\sigma^2H\ell_i\nu_j^c-\frac{1}{2}(M_R)_{ij}\nu^c_i\nu^c_j+\text{h.c.},
\end{equation}
where $\ell_i=(\nu_i,\:e_i)^T$, $e_i^c$, and $\nu^c_i$, are the
left-handed leptons, the right-handed charged lepton doublets, and the
right-handed SM singlet neutrinos, and $i=1,2,3$ is the generation
index. Here, $H$ is the SM Higgs doublet, $Y_\ell$ and $Y_D$ are the Dirac
Yukawa coupling matrices of the charged leptons and neutrinos, and $M_R$
is the Majorana mass matrix of the right-handed neutrinos with entries
of the order of the $B-L$ breaking scale $M_{B-L}\sim10^{14}\;\text{GeV}$. After electroweak symmetry breaking, $H$ develops a vacuum expectation value $\langle H\rangle\sim 10^2\:\text{GeV}$, and the mass terms of the leptons become
\begin{equation}\label{equ:massterms}
 \mathcal{L}_\text{mass}=-(M_\ell)_{ij}e_ie^c_j -(M_D)_{ij}\nu_i\nu_j^c-\frac{1}{2}(M_R)_{ij}\nu^c_i\nu^c_j+\text{h.c.},
\end{equation}
where $M_\ell=\langle H\rangle Y_\ell$ is the charged lepton and
$M_D=\langle H\rangle Y_D\sim 10^{2}\:\text{GeV}$ the Dirac neutrino
mass matrix. After integrating out the right-handed neutrinos, the
seesaw mechanism leads to the effective Majorana neutrino mass matrix
\begin{equation}
 M_\text{eff}=-M_DM_R^{-1}M_D^T~,\label{equ:Meff}
\end{equation}
with entries of the order $10^{-2}\;\text{eV}$ in agreement with
observation. The leptonic Dirac mass matrices $M_\ell$ and $M_D$, and the Majorana
mass matrices $M_R$ and $M_\text{eff}$ are diagonalized by
\begin{equation}
M_\ell=U_\ell M_\ell^\text{diag}U_{\ell'}^\dagger,\quad
M_D = U_D M_D^\text{diag} U_{D'}^\dagger,\quad M_R = U_R
M_R^\text{diag} U_R^T,\quad M_\text{eff}=U_\nu M_\text{eff}^\text{diag}U_\nu^T,
\label{equ:massmatrices}
\end{equation}
where $U_\ell,U_{\ell'},U_D,U_{D'},U_R$, and $U_\nu$, are unitary
mixing matrices,
whereas $M_\ell^\text{diag},M_D^\text{diag},M_R^\text{diag},$ and
$M_\text{eff}^\text{diag}$, are diagonal mass matrices with positive
entries. The mass eigenvalues of the charged leptons and neutrinos are given by
$M_\ell^\text{diag}=\text{diag}(m_e,m_\mu,m_\tau)$ and
$M^\text{diag}_\text{eff}=\text{diag}(m_1,m_2,m_3)$, where $m_1,m_2,$ and $m_3$,
are the first, second, and third neutrino mass eigenvalues. We can
always write a mixing matrix $U_x$ as a product of the form
\begin{equation}\label{equ:mixingmatrices}
 U_x=D_x\widehat{U}_xK_x,
\end{equation}
where $\widehat{U}_x$ is a CKM-like matrix that reads in the standard parameterization (we follow here throughout the
conventions and definitions given in \Ref~\cite{Plentinger:2006nb})
\begin{equation}
 \label{equ:ckm}
 \widehat{U}_x = \left( 
 \begin{array}{ccc}
   c_{12} c_{13} & s_{12} c_{13} & s_{13} e^{-\text{i}\widehat{\delta}^x} \\
   -s_{12} c_{23} - c_{12} s_{23} s_{13} e^{\text{i}\widehat{\delta}^x} &   c_{12} c_{23} -
 s_{12} s_{23} s_{13} e^{\text{i}\widehat{\delta}^x} & s_{23} c_{13} \\ 
 s_{12} s_{23} - c_{12} c_{23} s_{13}
e^{\text{i}\widehat{\delta}^x} & -c_{12} s_{23} - s_{12} c_{23} s_{13} e^{\text{i}\widehat{\delta}^x} & c_{23}
c_{13} 
 \end{array}
 \right),
\end{equation}
with $s_{ij} =\sin\hat{\theta}_{ij}^x$, $c_{ij} =
 \cos\hat{\theta}_{ij}^x$, and $\hat{\theta}_{ij}^x\in\{\hat{\theta}^x_{12},\hat{\theta}^x_{13},\hat{\theta}^x_{23}\}$ lie all in the first quadrant, i.e.,
$\hat{\theta}_{ij}\in[0,\frac{\pi}{2}]$, and 
$\widehat{\delta}^x\in[0,2\pi]$. In Eq.~(\ref{equ:mixingmatrices}),
$D_x$ and $K_x$ denote diagonal phase matrices that are $D_x=\text{diag}(e^{\text{i}\varphi_1^x},e^{\text{i}\varphi_2^x},e^{\text{i}\varphi_3^x})$
and
$K_x=\text{diag}(e^{\text{i}\alpha_1^x},e^{\text{i}\alpha_2^x},1)$,
where the index $x$ runs over $x=\ell,\ell',D,D',R,\nu$. The phases in
 $D_x$ and $K_x$ are all in the range $[0,2\pi]$.
The PMNS matrix reads
\begin{equation}\label{equ:pmnspara}
U_\text{PMNS}=U_\ell^\dagger U_\nu=\widehat{U}_\text{PMNS}K_\text{Maj},
\end{equation}
where $\widehat{U}$ is a CKM-like matrix parameterized as in
 \equ{ckm}, and
 $K_\text{Maj}=\text{diag}(e^{\text{i}\phi_1},e^{\text{i}\phi_2},1)$
 contains the Majorana phases $\phi_{1}$ and $\phi_2$. The CKM-like matrix $\widehat{U}_\text{PMNS}$ in \equ{pmnspara} is
described by the solar angle $\theta_{12}$, the reactor angle
$\theta_{13}$, the atmospheric angle $\theta_{23}$, and the Dirac
CP-phase $\delta$, which we identify in the parameterization of \equ{ckm} as
$\hat{\theta}_{ij}^x\rightarrow\theta_{ij}$ and
$\widehat{\delta}^x\rightarrow\delta$.

\section{Flavor Structure from $Z_n$ Symmetries}\label{sec:flavorsymmetries}
Let us next extend the SM gauge group to $G_\text{SM}\times G_F$,
where $G_F$ is a flavor symmetry. We assume
that $G_F$ is a direct product of discrete $Z_n$ symmetries, i.e.,
\begin{equation}\label{equ:productgroups}
G_F=Z_{n_1}\times Z_{n_2}\times\dots\times Z_{n_m},
\end{equation}
where $m$ is the number of $Z_n$ factors and the $n_k$
($k=1,2,\dots,m$) may be different. We will denote by $|G_F|=\Pi_{k=1}^mn_k$ the group order
(i.e., the number of elements) of $G_F$. Under $G_F$, we assign to the leptons the charges
\begin{equation}\label{equ:charges}
e^c_i\sim (p^i_1,p^i_2,\dots,p^i_m)=p^i,\quad
\ell_i\sim (q_1^i,q^i_2,\dots,q^i_m)= q^i,\quad
\nu^c_i\sim (r^i_1,r^i_2,\dots,r^i_m)= r^i,
\end{equation}
where the $j$th entry in each row vector denotes the
$Z_{n_j}$ charge of the particle and $i=1,2,3$ is the generation index
(see Sec.~\ref{sec:masses+mixings}). In the following, we choose a
convention, where for each group $Z_{n_k}$ the charges are non-negative
and lie in the range
\begin{equation}\label{equ:range}
p^i_k,q^i_k,r^i_k\in\{0,1,2,\dots,n_k-1\}.
\end{equation}
The sum $x_k+y_k$ of two $Z_{n_k}$ charges is equivalent with
 $x_k+y_k\;\text{mod}\;{n_k}$ and only determined modulo $n_k$. In
 order to spontaneously break the flavor symmetries, we assume for each factor
 $Z_{n_k}$ a single scalar flavon field $f_{n_k}$ that carries a
charge $-1\sim n_k-1$ under $Z_{n_k}$ but is a singlet under all other $Z_{n_j}$
with $j\neq k$. Moreover, the $f_{n_k}$ are $G_\text{SM}$ singlets.

When the $f_{n_k}$ acquire nonzero universal vacuum expectation values $\langle
f_{n_k}\rangle\simeq v$, non-renormalizable lepton Yukawa couplings and mass
terms of the form
\begin{eqnarray}\label{equ:nonrenormalizable}
 \mathcal{L}_\text{Y}&=&
-(\Pi_{k=1}^{m}\epsilon^{a^k_{ij}})
(Y'_\ell)_{ij}
H^\ast\ell_ie^c_j-
(\Pi_{k=1}^{m}\epsilon^{b^k_{ij}})(Y_D')_{ij}\text{i}\sigma^2H\ell_i\nu_j^c\nonumber\\
&&-\frac{1}{2}(\Pi_{k=1}^{m}\epsilon^{c^k_{ij}})
M_{B-L}(Y_R')_{ij}\nu^c_i\nu^c_j+\text{h.c.}
\end{eqnarray}
are generated by the Froggatt-Nielsen mechanism by integrating out
heavy fermions with universal mass $\simeq M_F$, where $\epsilon=v/M_F\simeq\theta_\text{C}\simeq0.2$ is of the order of
the Cabibbo angle, and 
\begin{eqnarray}\label{equ:exponents}
 a^k_{ij}&=&\text{min}\;\{p_i^k+q_j^k\;\text{mod}\,n_k,-p_i^k-q_j^k\;\text{mod}\,n_k\},\nonumber\\
 b^k_{ij}&=&\text{min}\;\{q_i^k+r_j^k\;\text{mod}\,n_k,-q_i^k-r_j^k\;\text{mod}\,n_k\},\\
 c^k_{ij}&=&\text{min}\;\{r_i^k+r_j^k\;\text{mod}\,n_k,
-r_i^k-r_j^k\;\text{mod}\,n_k\}\nonumber,
\end{eqnarray}
whereas $Y_\ell',Y'_D,$ and $Y_R'$, are dimensionless order
 unity Yukawa couplings. The modulo function $\text{mod}\,n_k$ in
 \equ{exponents} is a consequence of the cyclic nature of the $Z_n$ symmetries, and the
 minimum takes into account that the higher dimension operators
 can be built from both $f_{n_k}$ and the complex conjugated fields
 $f_{n_k}^\ast$. The important point is that the $Z_{n_k}$ charges of the
 leptons determine a hierarchical pattern of the Yukawa coupling matrices
 and the right-handed Majorana neutrino mass matrix.

 We define a ``texture'' as the
 matrix collecting the leading order products
 of $\epsilon$ for a certain Yukawa coupling or mass matrix, thereby
 ignoring the information on the order unity coefficients $Y_\ell',Y'_D,$ and $Y_R'$. The lepton textures are
 therefore the $3\times 3$ matrices with matrix elements approximating
 $(M_\ell)_{ij},(M_D)_{ij},$ and $(M_R)_{ij}$, as
\begin{equation}\label{equ:textures}
 (M_\ell)_{ij}\approx\Pi_{k=1}^{m}\epsilon^{a^k_{ij}},\quad
(M_D)_{ij}\approx\Pi_{k=1}^{m}\epsilon^{b^k_{ij}},\quad
(M_R)_{ij}\approx\Pi_{k=1}^{m}\epsilon^{c^k_{ij}}.
\end{equation}
In what follows, we will, for a certain model call the set of the three
 textures defined in Eq.~(\ref{equ:textures}) a ``texture set''.

Note that in our models, the $Z_n$ flavor symmetries are global, but it might be
important to gauge them to survive quantum gravity
corrections \cite{discreteanomalies}. The cancellation of anomalies
for our symmetries could, e.g.  be achieved by considering suitable extra matter
fields which is, however, beyond the
scope of this paper (for a recent related discussion of anomalies and
other phenomenology, see Ref.~\cite{Dreiner:2003yr}).
 
\section{Scanning Approach}\label{sec:selection}
In this section, we will describe how we identify among the models introduced in
Sec.~\ref{sec:flavorsymmetries} those which give nearly
tribimaximal lepton mixing. First, we pick some flavor symmetry group
$G_F$ as in \equ{productgroups} and assign to the three generations of leptons
$e^c_i,\ell_i,$ and $\nu^c_i$, all possible charge combinations under
$G_F$ according to Eqs.~(\ref{equ:charges}) and (\ref{equ:range}). Then, we determine from the charge assignments the
corresponding lepton textures for $M_\ell,M_D,$ and $M_R$, following
Eq.~(\ref{equ:textures}). Next, to find models
that can give a good fit to nearly tribimaximal lepton mixing, we
compare the textures found in Eq.~(\ref{equ:textures}) with the list of 1981
representative texture sets given in Ref.~\cite{Plentinger:2007px}. In
Ref.~\cite{Plentinger:2007px}, we have, based on assumptions of EQLC, determined fits of the order
one Yukawa couplings $Y_\ell',Y_D',$ and $Y_R'$, to reproduce nearly
tribimaximal lepton mixing along with a charged lepton mass
spectrum $m_e:m_\mu:m_\tau=\epsilon^4:\epsilon^2:1$ and a normal
neutrino mass hierarchy $m_1:m_2:m_3=\epsilon^2:\epsilon:1$ in
perfect agreement with current neutrino data (at the $3\sigma$
confidence level (CL)).
 
In Ref.~\cite{Plentinger:2007px}, the textures have been extracted in the
basis where $U_{\ell'}=\mathbbm{1}$, i.e., where the rotations acting on the $e^c_i$ are zero. Although these rotations
do not show up in the observables, they are important for formulating
our explicit models using flavor symmetries. We therefore include now
in our considerations textures of the charged leptons that have been extracted in bases where
$U_{\ell'}$ can be nontrivial, i.e., where we can have $U_{\ell'}\neq\mathbbm{1}$, thereby
leading to a multitude of new explicit representations for the charged
lepton textures. In complete analogy with
Ref.~\cite{Plentinger:2007px}, we assume for the mixing angles and
phases entering the matrices $U_{\ell'}$ all possible combinations that satisfy
\begin{equation}
\theta^{\ell'}_{12},\theta^{\ell'}_{13},\theta^{\ell'}_{23}\in\{0,\epsilon^2,\epsilon,\frac{\pi}{4}\},
\end{equation}
while the phases can take the values
$\widehat{\delta}^{\ell'},\varphi^{\ell'}_1,\varphi^{\ell'}_2\in\{0,\pi\}$,
whereas we can always choose $\varphi_3^{\ell'}=\alpha^{\ell'}_1=\alpha_2^{\ell'}=0$ (for a
definition of the notation, see Sec.~\ref{sec:masses+mixings}). This
applies the hypothesis of EQLC to the $e^c_i$ such that the left- and right-handed lepton sectors
are now all treated on the same footing. As a consequence, after
employing the texture reduction from
Refs.~\cite{Plentinger:2006nb,Plentinger:2007px}, we finally arrive at an enlarged ``reference list'' of 43278
qualitatively different lepton texture sets. We use this list as our reference
to match onto flavor symmetry models for nearly tribimaximal lepton mixing. Any flavor charge
assignment that yields textures contained in this reference list
provides a valid model that allows an excellent fit to nearly
tribimaximal lepton mixing.\footnote{Note that in our reference list (after
  factoring out a possible overall power of $\epsilon$) we set an entry $\epsilon^n$ equal to zero when $n\geq
3$ (for neutrinos only) or when $n\geq 5$ (for charged leptons). This
is different from Ref.~\cite{Plentinger:2007px}, where such an entry
is set to zero when $n\geq 3$ for both neutrinos and charged
leptons. For specific flavor models, however, we will always show the
actual suppression factor $\epsilon^n$, as predicted by the flavor
symmetry.} 

We can now impose extra assumptions on the properties of the textures
in our reference list to search for interesting flavor symmetry models. For example,
we will demand that none of the textures is completely anarchic (or
democratic) and that the charged lepton textures contain (after
factoring out common factors) at least one
entry $\epsilon^n$ with $n\geq4$ (to have sufficient structure in the texture). This reduces the above
reference list to a reduced list of 17772 distinct texture sets. In the next section, we
will use this reduced reference list to perform the group space scan of flavor symmetries.

\section{Results of the Group Space Scan}\label{sec:scan}
Let us now present the results of the group space scan for flavor
symmetries that produce nearly tribimaximal lepton mixing. We assume
throughout the models introduced in Sec.~\ref{sec:flavorsymmetries}
and choose as flavor symmetries\footnote{For an application of related flavor
  groups see Ref.~\cite{Sayre:2007ps}.} $G_F=\Pi_{k=1}^m Z_{n_k}$ for
$m=1,2,3,4$. A complete scan has been performed for groups up to order
40 (for $m=1$), 45 (for $m=2$), 30 (for $m=3$), and 24 (for $m=4$),
with $n_k\leq 9$ for $m>1$. Valid models are selected as described in Sec.~\ref{sec:selection} by matching the
textures generated by the flavor charge assignments onto the reduced
reference list of 17772 non-anarchic texture sets.
\begin{figure}[th]
\begin{center}
\includegraphics*[bb=470 40 0 370, height=12cm]{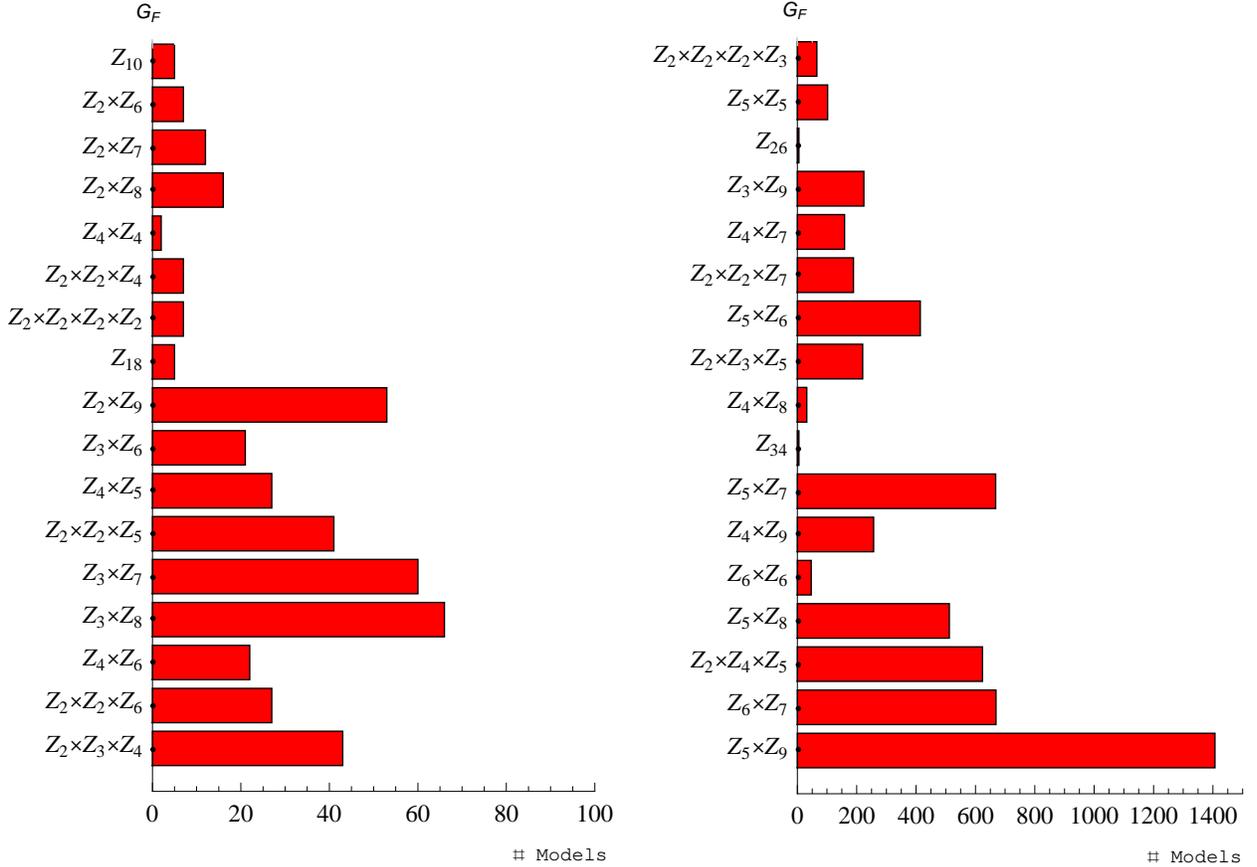}
\caption{\label{fig:allmodels} Number of flavor models leading to nearly
  tribimaximal lepton mixing as a function of the flavor
  group $G_F$ for increasing group order. In the left (right) panel we have
  $10\leq|G_F|\leq 24$ ($24\leq|G_F|\leq 45$).}
\end{center}
\end{figure}
In total, we find in the scan 6021 such models that reproduce 2093
texture sets. The distribution of the models for the different groups $G_F$
  is summarized in \figu{allmodels}.\footnote{Additionally, we have also
  included the number of valid models for $G_F=Z_2\times Z_4\times
  Z_5$.} All these models allow for an excellent fit 
to nearly tribimaximal neutrino mixing (at $3\sigma$ CL and
most of them actually at $1\sigma$ CL \cite{Plentinger:2007px}) with a
very small reactor angle $\theta_{13}\lesssim 1^\circ$ and the lepton
mass ratios\footnote{We are interested here in an $SU(5)$ compatible fit.}
\begin{equation}\label{equ:massratios}
 m_e:m_\mu:m_\tau=\epsilon^4:\epsilon^2:1,\quad m_1:m_2:m_3=\epsilon^2:\epsilon:1,
\end{equation}
i.e., we have have a normal neutrino mass hierarchy
with a ratio $\Delta m_\odot^2/\Delta m_\text{atm}^2\sim\epsilon^2$ of
solar over atmospheric neutrino mass squared difference. Since the
neutrinos have a normal mass hierarchy, renormalization group effects
are negligible (for a discussion and references see
Ref.~\cite{Plentinger:2007px}).

In \figu{allmodels}, we can observe the trend that the number of
 valid models generally increases with the group order $|G_F|$ up to some periodic
 modulation. It is interesting to give a rough estimate on how large $G_F$
 has to be in order to more or less reproduce an arbitrary texture
 set. For this purpose, note that (after factoring out a possible common overall factor
 $\epsilon^n$) we restrict ourselves in $M_D$ and $M_R$ to the matrix entries
 $\{0,\epsilon^2,\epsilon,1\}$, and in $M_\ell$ to the entries $\{0,\epsilon^4,\epsilon^3,\epsilon^2,\epsilon,1\}$, i.e., 
we have $4^{6+9}\cdot 6^9$ different possibilities for arbitrary
 texture sets.
On the other hand, we have nine different charges per group, leading
 to $|G_F|^9$
different possibilities for the charge assignments in $G_F$ (cf.~\equ{charges}). Note that
 we use $|G_F|$ as a figure of merit: The larger $|G_F|$, the more
 possibilities for the charge assignments we have. In order to
 reproduce {\em any} texture set, we roughly estimate that the
number of possibilities for the charges should exceed the number of
 possibilities for the texture sets, i.e.,
\begin{equation}
|G_F|\gtrsim 4^{\frac{15}{9}}\cdot 6 \simeq 60.
\label{equ:est}
\end{equation}
This means that, for instance, four $Z_n$ factors with moderate $n_k$, such
as $Z_2\times Z_3\times Z_4\times Z_5$ should be sufficient, or two
$Z_n$ factors with large enough $n_k$, such as $Z_7$ and higher. In \figu{allmodels}, we
have much less possibilities on the l.h.s. where $10\leq
|G_F|\leq 24$, while we have on the r.h.s. $24\leq|G_F|\leq 45$. In
fact, \figu{allmodels} seems to suggest that $G_F=Z_5\times Z_9$ with
$|G_F|=45$ is already entering the regime estimated in \equ{est}.

We have checked that for the 6021 valid models practically all (i.e., more
 than 99\%) of the Yukawa
 coupling matrix elements of $Y_\ell',Y_D',$ and $Y_R'$, (see \equ{nonrenormalizable}) lie in the interval between $\epsilon$
 and $1/\epsilon$. With respect to the expansion
 parameter $\epsilon$ of our models, these matrix elements can therefore indeed be viewed as order one coefficients.

Out of our set of 6021 valid models, let us now consider a few
examples. In Tab.~\ref{tab:models}, we show 22 explicit valid models by listing the flavor group with complete flavor charge assignment and the resulting textures for $M_\ell,M_D,$ and
$M_R$. The rough guideline for choosing these 22 models was to give one
example for each texture set previously identified in EQLC
\cite{Plentinger:2007px}, with distinct $M_R$, a charged lepton mass
spectrum as in \equ{massratios}, and ``natural'' order one Yukawa
couplings $Y'_\ell, Y_D',$ and $Y_R'$. The complete information on the corresponding mass and mixing parameters of the 22
models is summarized in \Tab~\ref{tab:modeldata} in the Appendix.

\Tab~\ref{tab:modeldata} demonstrates that our models are very
 general in the sense that they can exhibit maximal mixings in the charged lepton and/or the neutrino
 sector, and that the left- and/or right-handed neutrinos can mix maximally. In addition,
 we can have maximal mixings in all sectors, not only between the 2nd
 and 3rd generation, but also between the 1st and the
 2nd as well as between the 1st and the 3rd generation.

All 22 models in \Tab s \ref{tab:models} and \ref{tab:modeldata}
lead to the following PMNS mixing angles in the range
\begin{equation}
34^\circ\lesssim\theta_{12}\lesssim 39^\circ,\quad\theta_{13}\lesssim
 1^\circ,\quad \theta_{23}\approx 52^\circ,
\end{equation}
in agreement with neutrino oscillation data (at $3\sigma$ CL). The models are thus characterized by a very small
reactor angle close to zero, and a significant deviation of about $+7^\circ$ from maximal
atmospheric mixing.
\setlength{\LTcapwidth}{\textwidth}
\begin{longtable}{@{\hspace{0mm}}c@{\hspace{0mm}}|c|c|c|@{\hspace{0mm}}c@{\hspace{0mm}}|@{\hspace{1mm}}c@{\hspace{0mm}}}
\#\hspace{2mm} & $M_\ell/\langle H\rangle$ & $M_D/\langle H\rangle$ & $M_R/M_{B-L}$ & $
\ba{c} p^1,p^2,p^3\\q^1,q^2,q^3\\r^1,r^2,r^3\ea$&$G_F$ \\&&&&&\\[-3.5mm]\hline&&&&&\\[-3mm]
\endfirsthead
1 & $\left(\ba{ccc}\epsilon^{4}&\epsilon^{5}&\epsilon^{2}\\ \epsilon^{2}&\epsilon^{2}&\epsilon^{2}\\ \epsilon^{2}&\epsilon^{4} & 1 \ea\right)$ &  
$\epsilon\left(\ba{ccc}\epsilon & \epsilon^{2} & \epsilon^{2}\\ \epsilon & 1 & \epsilon \\ \epsilon &  1 & \epsilon \ea\hspace{-1mm}\right)$ &
$\epsilon^{3}\hspace{-1mm}
\left(\hspace{-1mm}\ba{ccc}
1 & \epsilon^{2} & 1 \\
\epsilon^{2} & 1 & \epsilon^{2} \\
1 & \epsilon^{2} & 1 \ea\hspace{-1mm}\right)$ & $\ba{c}(2,0),(0,0),(2,5)\\(2,3),(4,1),(3,2)\\(1,4),(2,6),(0,5)\ea$
& $Z_5\times Z_7$ \\[6.5mm]\hline&&&&&\\*[-3mm]

2 & $\epsilon
\hspace{-1mm}\left(\hspace{-1mm}\ba{ccc}
\epsilon^{4}&\epsilon^{4}&\epsilon^{2} \\ \epsilon^{3}&\epsilon^{2} & 1 \\ \epsilon^{3}&\epsilon^{4} & 1 \ea\hspace{-1mm}\right)$ &
$\epsilon
\hspace{-1mm}\left(\hspace{-1mm}\ba{ccc}
\epsilon &\epsilon^{3}&\epsilon \\ \epsilon &  1 & \epsilon^{3} \\ \epsilon &\epsilon^{2}&\epsilon \ea\hspace{-1mm}\right)$ &
$\epsilon^{2}\hspace{-1mm}
\left(\hspace{-1mm}\ba{ccc}
\epsilon &\epsilon &\epsilon \\
\epsilon &  1 & \epsilon^{2} \\
\epsilon &\epsilon^{2} & 1 \ea\hspace{-1mm}\right)$ & $\ba{c}(2,2),(3,2),(2,5)\\(0,1),(2,2),(4,2)\\(2,6),(3,4),(1,0)\ea$
& $Z_{5}\times Z_{7}$ \\[6.5mm]\hline&&&&&\\*[-3mm]

3 & $\epsilon
\hspace{-1mm}\left(\hspace{-1mm}\ba{ccc}
\epsilon^{4}&\epsilon^{3}&\epsilon^{5} \\ \epsilon^{3}&\epsilon^{2}&\epsilon^{2} \\ \epsilon &\epsilon^{2} & 1 \ea\hspace{-1mm}\right)$ &
$\epsilon^{2}
\hspace{-1mm}\left(\hspace{-1mm}\ba{ccc}
\epsilon &\epsilon &\epsilon^{3} \\ \epsilon &  1 & 1 \\ \epsilon &  1 & 1 \ea\hspace{-1mm}\right)$ &
$\epsilon
\hspace{-1mm}\left(\hspace{-1mm}\ba{ccc}
\epsilon &\epsilon &\epsilon^{5} \\ \epsilon &  1 & \epsilon^{4} \\ \epsilon^{5}&\epsilon^{4} & 1 \ea\hspace{-1mm}\right)$ & $\ba{c}
(3,7),(3,0),(2,7)\\(1,5),(3,6),(3,2)\\(1,4),(2,4),(2,0)\ea$
& $Z_{5}\times Z_{8}$ \\[6.5mm]\hline&&&&&\\*[-3mm]

4 & $\epsilon
\hspace{-1mm}\left(\hspace{-1mm}\ba{ccc}
\epsilon^{3}&\epsilon^{3}&\epsilon \\ \epsilon^{2}&\epsilon^{2}&\epsilon^{2} \\ \epsilon^{2}&\epsilon^{4} & 1 \ea\hspace{-1mm}\right)$ &
$\epsilon^{3}
\hspace{-1mm}\left(\hspace{-1mm}\ba{ccc}
\epsilon &\epsilon &\epsilon \\ \epsilon &  1 & 1 \\ \epsilon &  1 & 1 \ea\hspace{-1mm}\right)$ &
$\epsilon
\hspace{-1mm}\left(\hspace{-1mm}\ba{ccc}
\epsilon &\epsilon^{5}&\epsilon \\ \epsilon^{5} & 1 & \epsilon^{4} \\ \epsilon &\epsilon^{4} & 1 \ea\hspace{-1mm}\right)$ & $\ba{c}
(3,0),(0,1),(2,5)\\(4,2),(3,6),(3,2)\\(4,0),(3,4),(3,0)\ea$
& $Z_{5}\times Z_{8}$ \\[6.5mm]\hline&&&&&\\*[-3mm]

5 & $\epsilon
\hspace{-1mm}\left(\hspace{-1mm}\ba{ccc}
\epsilon^{4}&\epsilon^{2}&\epsilon \\ \epsilon^{3}&\epsilon^{2}&\epsilon^{2} \\ \epsilon^{5}&\epsilon &  1 \ea\hspace{-1mm}\right)$ &
$\epsilon^{2}
\hspace{-1mm}\left(\hspace{-1mm}\ba{ccc}
\epsilon &\epsilon &\epsilon^{3} \\ \epsilon^{3} & 1 & \epsilon \\ \epsilon^{3} & 1 & \epsilon \ea\hspace{-1mm}\right)$ &
$\epsilon^{2}
\hspace{-1mm}\left(\hspace{-1mm}\ba{ccc}
1&\epsilon^{3} & 1 \\ \epsilon^{3} & 1 & \epsilon^{4} \\ 1&\epsilon^{4} & 1 \ea\hspace{-1mm}\right)$ & $\ba{c}
(3,8),(4,3),(0,3)\\(0,4),(3,7),(4,6)\\(0,8),(2,4),(1,0)\ea$
& $Z_{5}\times Z_{9}$ \\[6.5mm]\hline&&&&&\\*[-3mm]

6 & $\epsilon
\hspace{-1mm}\left(\hspace{-1mm}\ba{ccc}
\epsilon^{4}&\epsilon^{2}&\epsilon \\ \epsilon^{3}&\epsilon^{2}&\epsilon^{2} \\ \epsilon^{5}&\epsilon &  1 \ea\hspace{-1mm}\right)$ &
$\epsilon^{2}
\hspace{-1mm}\left(\hspace{-1mm}\ba{ccc}
\epsilon^{2}&\epsilon^{3}&\epsilon \\ \epsilon &  1 & 1 \\ \epsilon &  1 & 1 \ea\hspace{-1mm}\right)$ &
$\epsilon^{2}
\hspace{-1mm}\left(\hspace{-1mm}\ba{ccc}
\epsilon^{2}&\epsilon &\epsilon \\ \epsilon &  1 & 1 \\ \epsilon &  1 & 1 \ea\hspace{-1mm}\right)$ & $\ba{c}
(3,6),(2,1),(1,1)\\(4,6),(1,0),(0,8)\\(1,8),(0,8),(1,0)\ea$
& $Z_{5}\times Z_{9}$ \\[6.5mm]\hline&&&&&\\*[-3mm]

7 & $\epsilon
\hspace{-1mm}\left(\hspace{-1mm}\ba{ccc}
\epsilon^{4}&\epsilon^{4}&\epsilon^{2} \\ \epsilon^{3}&\epsilon^{2} & 1 \\ \epsilon^{3}&\epsilon^{4} & 1 \ea\hspace{-1mm}\right)$ &
$\epsilon^{4}
\hspace{-1mm}\left(\hspace{-1mm}\ba{ccc}
\epsilon^{2}&\epsilon &\epsilon^{2} \\ \epsilon &\epsilon &\epsilon \\ 1&\epsilon &  1 \ea\hspace{-1mm}\right)$ &
$\epsilon
\hspace{-1mm}\left(\hspace{-1mm}\ba{ccc}
1&\epsilon^{2} & 1 \\ \epsilon^{2}&\epsilon &\epsilon^{2} \\ 1&\epsilon^{2} & 1 \ea\hspace{-1mm}\right)$ & $\ba{c}
(2,8),(1,8),(1,4)\\(1,4),(4,4),(3,5)\\(2,0),(0,1),(2,0)\ea$
& $Z_{5}\times Z_{9}$ \\[6.5mm]\hline&&&&&\\*[-3mm]

8 & $\epsilon
\hspace{-1mm}\left(\hspace{-1mm}\ba{ccc}
\epsilon^{3}&\epsilon^{4}&\epsilon \\ \epsilon^{3}&\epsilon^{2}&\epsilon^{2} \\ \epsilon^{2}&\epsilon^{4} & 1 \ea\hspace{-1mm}\right)$ &
$\epsilon^{2}
\hspace{-1mm}\left(\hspace{-1mm}\ba{ccc}
\epsilon^{2}&\epsilon &\epsilon^{3} \\ \epsilon^{2} & 1 & 1 \\ \epsilon^{2} & 1 & 1 \ea\hspace{-1mm}\right)$ &
$\epsilon^{2}
\hspace{-1mm}\left(\hspace{-1mm}\ba{ccc}
\epsilon^{2}&\epsilon^{2}&\epsilon^{2} \\ \epsilon^{2} & 1 & 1 \\ \epsilon^{2} & 1 & 1 \ea\hspace{-1mm}\right)$ & $\ba{c}
(2,1),(1,6),(4,1)\\(1,6),(0,1),(1,0)\\(3,6),(0,1),(1,0)\ea$
& $Z_{5}\times Z_{9}$ \\[6.5mm]\hline&&&&&\\*[-3mm]

9 & \hspace{-1mm}$\left(\hspace{-1mm}\ba{ccc}
\epsilon^{4}&\epsilon^{3}&\epsilon^{5} \\ \epsilon^{3}&\epsilon^{2}&\epsilon^{2} \\ \epsilon &\epsilon^{4} & 1 \ea\hspace{-1mm}\right)$ &
$\epsilon^{3}
\hspace{-1mm}\left(\hspace{-1mm}\ba{ccc}
\epsilon^{2}&\epsilon &\epsilon \\ 1 & 1 & \epsilon^{2} \\ 1 & 1 & \epsilon \ea\hspace{-1mm}\right)$ & $\epsilon^{2}
\hspace{-1mm}\left(\hspace{-1mm}\ba{ccc}
1 & 1 & \epsilon \\ 1 & 1 & \epsilon \\ \epsilon &\epsilon &  1 \ea\hspace{-1mm}\right)$
& $\ba{c}
(0,2),(2,5),(1,2)\\(2,3),(4,4),(5,5)\\(2,0),(3,1),(0,6)\ea$
& $Z_{6}\times Z_{7}$ \\[6.5mm]\hline&&&&&\\*[-3mm]

10 & \hspace{-1mm}$\left(\hspace{-1mm}\ba{ccc}
\epsilon^{4}&\epsilon^{6}&\epsilon^{2} \\ \epsilon^{2}&\epsilon^{2}&\epsilon^{2} \\ \epsilon^{2}&\epsilon^{4} & 1 \ea\hspace{-1mm}\right)$ &
$\epsilon^{2}
\hspace{-1mm}\left(\hspace{-1mm}\ba{ccc}
\epsilon^{2}&\epsilon &\epsilon^{2} \\ \epsilon^{2}&\epsilon &  1 \\ \epsilon^{2}&\epsilon &  1 \ea\hspace{-1mm}\right)$ & $\epsilon
\hspace{-1mm}\left(\hspace{-1mm}\ba{ccc}
\epsilon^{2}&\epsilon^{2}&\epsilon^{2} \\ \epsilon^{2}&\epsilon &\epsilon^{4} \\ \epsilon^{2}&\epsilon^{4} & 1 \ea\hspace{-1mm}\right)$
& $\ba{c}
(2,1),(2,3),(0,1)\\(1,0),(5,5),(0,6)\\(4,3),(2,0),(0,3)\ea$
& $Z_{6}\times Z_{7}$ \\[6.5mm]\hline&&&&&\\*[-3mm]

11 & \hspace{-1mm}$\left(\hspace{-1mm}\ba{ccc}
\epsilon^{3}&\epsilon^{3}&\epsilon^{2} \\ \epsilon^{2}&\epsilon^{2}&\epsilon^{2} \\ \epsilon^{4}&\epsilon^{2} & 1 \ea\hspace{-1mm}\right)$ & $
\epsilon^{2}
\hspace{-1mm}\left(\hspace{-1mm}\ba{ccc}
\epsilon &\epsilon &\epsilon \\ \epsilon &  1 & \epsilon \\ \epsilon &  1 & \epsilon \ea\hspace{-1mm}\right)$ & $\epsilon
\hspace{-1mm}\left(\hspace{-1mm}\ba{ccc}
\epsilon &\epsilon &\epsilon^{2} \\ \epsilon &  1 & \epsilon^{3} \\ \epsilon^{2}&\epsilon^{3} & 1 \ea\hspace{-1mm}\right)$
& $\ba{c}
(0,0,0),(1,0,2),(1,1,3)\\(1,1,1),(0,0,2),(1,2,2)\\(1,0,1),(1,0,2),(0,1,0)\ea$
& $Z_{2}\times Z_{3}\times Z_{5}$ \\[6.5mm]\hline&&&&&\\*[-3mm]

12 & \hspace{-1mm}$\left(\hspace{-1mm}\ba{ccc}
\epsilon^{4}&\epsilon^{5}&\epsilon \\ \epsilon^{3}&\epsilon^{2}&\epsilon^{2} \\ \epsilon^{5}&\epsilon^{4} & 1 \ea\hspace{-1mm}\right)$ &
$\epsilon^{2}
\hspace{-1mm}\left(\hspace{-1mm}\ba{ccc}
\epsilon &\epsilon &\epsilon^{2} \\ \epsilon^{2} & 1 & \epsilon^{2} \\ \epsilon^{2} & 1 & \epsilon \ea\hspace{-1mm}\right)$ & $\epsilon^{2}$
\hspace{-1mm}$\left(\hspace{-1mm}\ba{ccc}
1&\epsilon &  1 \\ \epsilon &  1 & \epsilon \\ 1&\epsilon &  1 \ea\hspace{-1mm}\right)$
& $\ba{c}
(1,2,4),(1,1,4),(0,0,2)\\(0,1,3),(1,0,2),(0,0,3)\\(0,2,4),(1,0,1),(0,1,0)\ea$
& $Z_{2}\times Z_{4}\times Z_{5}$ \\[6.5mm]\hline&&&&&\\*[-3mm]

13 & \hspace{-1mm}$\left(\hspace{-1mm}\ba{ccc}
\epsilon^{4}&\epsilon^{4}&\epsilon^{2} \\ \epsilon^{3}&\epsilon^{2}&\epsilon^{2} \\ \epsilon^{5}&\epsilon^{4} & 1 \ea\hspace{-1mm}\right)$ & $
\epsilon^{2}
\hspace{-1mm}\left(\hspace{-1mm}\ba{ccc}
\epsilon^{3}&\epsilon &\epsilon^{2} \\ \epsilon^{2}&\epsilon &  1 \\ \epsilon^{2}&\epsilon &  1 \ea\hspace{-1mm}\right)$ & $
\hspace{-1mm}\left(\hspace{-1mm}\ba{ccc}
\epsilon^{2}&\epsilon^{3}&\epsilon^{2} \\ \epsilon^{3}&\epsilon &\epsilon^{2} \\ \epsilon^{2}&\epsilon^{2} & 1 \ea\hspace{-1mm}\right)$
& $\ba{c}
(1,1,1),(1,1,0),(0,3,4)\\(0,2,2),(0,3,1),(0,1,1)\\(1,0,1),(0,0,2),(0,0,0)\ea$
& $Z_{2}\times Z_{4}\times Z_{5} $\\[6.5mm]\hline&&&&&\\*[-3mm]

14 & $
\epsilon
\hspace{-1mm}\left(\hspace{-1mm}\ba{ccc}
\epsilon^{4}&\epsilon^{5}&\epsilon^{2} \\ \epsilon^{3}&\epsilon^{2} & 1 \\ \epsilon^{3}&\epsilon^{4} & 1 \ea\hspace{-1mm}\right)$ & $
\epsilon
\hspace{-1mm}\left(\hspace{-1mm}\ba{ccc}
\epsilon^{2}&\epsilon &\epsilon^{2} \\ \epsilon^{2}&\epsilon &\epsilon^{2} \\ 1&\epsilon^{3} & 1 \ea\hspace{-1mm}\right)$ & $\epsilon^{3}
\hspace{-1mm}\left(\hspace{-1mm}\ba{ccc}
1&\epsilon &  1 \\ \epsilon &\epsilon &\epsilon \\ 1&\epsilon &  1 \ea\hspace{-1mm}\right)$ & $\ba{c}
(0,3),(3,3),(0,6)\\(3,1),(1,3),(3,3)\\(1,5),(3,8),(1,5)\ea$
& $Z_{4}\times Z_{9}$\\[6.5mm]\hline&&&&&\\*[-3mm]

15 & $
\epsilon
\hspace{-1mm}\left(\hspace{-1mm}\ba{ccc}
\epsilon^{4}&\epsilon^{4}&\epsilon^{2} \\ \epsilon^{3}&\epsilon^{4} & 1 \\ \epsilon^{3}&\epsilon^{2} & 1 \ea\hspace{-1mm}\right)$ &$
\epsilon^{2}
\hspace{-1mm}\left(\hspace{-1mm}\ba{ccc}
\epsilon^{2}&\epsilon &\epsilon^{3} \\ \epsilon^{2}&\epsilon &\epsilon^{2} \\ \epsilon &\epsilon^{2} & 1 \ea\hspace{-1mm}\right)$ & $\epsilon
\hspace{-1mm}\left(\hspace{-1mm}\ba{ccc}
\epsilon^{2}&\epsilon^{2}&\epsilon \\ \epsilon^{2}&\epsilon &\epsilon^{2} \\ \epsilon &\epsilon^{2} & 1 \ea\hspace{-1mm}\right)$ & $\ba{c}
(4,2),(0,2),(4,5)\\(3,1),(2,2),(0,2)\\(1,3),(2,3),(0,3)\ea$
& $Z_{5}\times Z_{7}$ \\[6.5mm]\hline&&&&&\\*[-3mm]

16 & $
\epsilon^2
\hspace{-1mm}\left(\hspace{-1mm}\ba{ccc}
\epsilon^{4}&\epsilon^{3}&\epsilon^{2} \\ \epsilon^{3}&\epsilon^{2} & 1 \\ \epsilon^{3}&\epsilon^{4} & 1 \ea\hspace{-1mm}\right)$ & $
\epsilon^2
\hspace{-1mm}\left(\hspace{-1mm}\ba{ccc}
\epsilon^3 &\epsilon &\epsilon \\ 1 & 1 & \epsilon^3 \\ \epsilon^2 &\epsilon &\epsilon \ea\hspace{-1mm}\right)$ & $\epsilon^{2}
\hspace{-1mm}\left(\hspace{-1mm}\ba{ccc}
1 &1 &\epsilon^3 \\ 1 &  1 & \epsilon^3 \\ \epsilon^3 &  \epsilon^3 & 1 \ea\hspace{-1mm}\right)$ & $\ba{c}
(1,6),(0,5),(1,0)\\(2,7),(0,8),(3,8)\\(0,8),(4,0),(2,4)\ea$
& $Z_{5}\times Z_{9}$ \\[6.5mm]\hline&&&&&\\*[-3mm]

17 & $
\epsilon
\hspace{-1mm}\left(\hspace{-1mm}\ba{ccc}
\epsilon^{4}&\epsilon^{4}&\epsilon^{2} \\ \epsilon^{3}&\epsilon^{4} & 1 \\ \epsilon^{3}&\epsilon^{2} & 1 \ea\hspace{-1mm}\right)$ & $
\epsilon
\hspace{-1mm}\left(\hspace{-1mm}\ba{ccc}
\epsilon^{2}&\epsilon &\epsilon^{3} \\ \epsilon^{2}&\epsilon &\epsilon^{2} \\ \epsilon^{3}&\epsilon^{2} & 1 \ea\hspace{-1mm}\right)$ & $\epsilon^{2}
\hspace{-1mm}\left(\hspace{-1mm}\ba{ccc}
\epsilon^{2}&\epsilon^{2}&\epsilon^{3} \\ \epsilon^{2}&\epsilon &\epsilon^{2} \\ \epsilon^{3}&\epsilon^{2} & 1 \ea\hspace{-1mm}\right)$ & $\ba{c}
(4,3),(0,3),(4,0)\\(3,1),(2,0),(0,0)\\(2,2),(2,1),(0,1)\ea$
& $Z_{5}\times Z_{7}$ \\[6.5mm]\hline&&&&&\\*[-3mm]

18 & $
\hspace{-1mm}\left(\hspace{-1mm}\ba{ccc}
\epsilon^{4}&\epsilon^{4}&\epsilon^{2} \\ \epsilon^{3}&\epsilon^{2}&\epsilon^{2} \\ \epsilon^{5}&\epsilon^{2} & 1 \ea\hspace{-1mm}\right)$ & $
\epsilon^{2}
\hspace{-1mm}\left(\hspace{-1mm}\ba{ccc}
\epsilon^{4}&\epsilon &\epsilon^{2} \\ \epsilon^{2}&\epsilon &  1 \\ \epsilon^{2}&\epsilon &  1 \ea\hspace{-1mm}\right)$ & $\epsilon
\hspace{-1mm}\left(\hspace{-1mm}\ba{ccc}
\epsilon^{2}&\epsilon^{3}&\epsilon^{2} \\ \epsilon^{3}&\epsilon &\epsilon^{5} \\ \epsilon^{2}&\epsilon^{5} & 1 \ea\hspace{-1mm}\right)$ & $\ba{c}
(4,4),(1,2),(0,1)\\(1,0),(1,5),(0,6)\\(2,3),(3,1),(0,3)\ea$
& $Z_{6}\times Z_{7}$ \\[6.5mm]\hline&&&&&\\*[-3mm]

19 & $
\hspace{-1mm}\left(\hspace{-1mm}\ba{ccc}
\epsilon^{4}&\epsilon^{4}&\epsilon^{2} \\ \epsilon^{2}&\epsilon^{2}&\epsilon^{2} \\ \epsilon^{4}&\epsilon^{2} & 1 \ea\hspace{-1mm}\right)$ & $
\epsilon^{2}
\hspace{-1mm}\left(\hspace{-1mm}\ba{ccc}
\epsilon &\epsilon^{2}&\epsilon \\ \epsilon &  1 & \epsilon \\ \epsilon &  1 & \epsilon \ea\hspace{-1mm}\right)$ & $
\hspace{-1mm}\left(\hspace{-1mm}\ba{ccc}
\epsilon &\epsilon^{2}&\epsilon^{5} \\ \epsilon^{2} & 1 & \epsilon^{3} \\ \epsilon^{5}&\epsilon^{3} & 1 \ea\hspace{-1mm}\right)$ & $\ba{c}
(0,1),(0,3),(4,4)\\(2,1),(1,4),(1,2)\\(2,3),(0,3),(0,0)\ea$
& $Z_{5}\times Z_{6}$ \\[6.5mm]\hline&&&&&\\*[-3mm]

20 & $
\hspace{-1mm}\left(\hspace{-1mm}\ba{ccc}
\epsilon^{4}&\epsilon^{5}&\epsilon^{2} \\ \epsilon^{2}&\epsilon^{2}&\epsilon^{2} \\ \epsilon^{4}&\epsilon^{4} & 1 \ea\hspace{-1mm}\right)$ & $
\epsilon
\hspace{-1mm}\left(\hspace{-1mm}\ba{ccc}
\epsilon^{2}&\epsilon &\epsilon^{2} \\ \epsilon^{2}&\epsilon &  1 \\ \epsilon^{2}&\epsilon &  1 \ea\hspace{-1mm}\right)$ & $\epsilon
\hspace{-1mm}\left(\hspace{-1mm}\ba{ccc}
\epsilon^{2}&\epsilon^{2}&\epsilon^{2} \\ \epsilon^{2}&\epsilon &\epsilon^{2} \\ \epsilon^{2}&\epsilon^{2} & 1 \ea\hspace{-1mm}\right)$ & $\ba{c}
(2,0),(3,5),(1,3)\\(0,4),(3,2),(4,3)\\(2,1),(0,4),(2,3)\ea$
&$ Z_{5}\times Z_{6}$ \\[6.5mm]\hline&&&&&\\*[-3mm]

21 & $
\hspace{-1mm}\left(\hspace{-1mm}\ba{ccc}
\epsilon^{4}&\epsilon^{5}&\epsilon^{2} \\ \epsilon^{2}&\epsilon^{2}&\epsilon^{2} \\ \epsilon^{4}&\epsilon^{4} & 1 \ea\hspace{-1mm}\right)$ & $
\epsilon
\hspace{-1mm}\left(\hspace{-1mm}\ba{ccc}
\epsilon &\epsilon^{2}&\epsilon \\ \epsilon &  1 & \epsilon \\ \epsilon^{3} & 1 & \epsilon \ea\hspace{-1mm}\right)$ &$ \epsilon^{2}
\hspace{-1mm}\left(\hspace{-1mm}\ba{ccc}
\epsilon &\epsilon^{2}&\epsilon \\ \epsilon^{2} & 1 & \epsilon^{3} \\ \epsilon &\epsilon^{3} & 1 \ea\hspace{-1mm}\right)$ & $\ba{c}
(3,4),(4,4),(1,2)\\(3,5),(4,2),(4,4)\\(2,5),(1,3),(1,0)\ea$
& $Z_{5}\times Z_{6}$ \\[6.5mm]\hline&&&&&\\*[-3mm]

22 & $
\hspace{-1mm}\left(\hspace{-1mm}\ba{ccc}
\epsilon^{4}&\epsilon^{3}&\epsilon^{2} \\ \epsilon^{2}&\epsilon^{2}&\epsilon^{3} \\ \epsilon^{5}&\epsilon &  1 \ea\hspace{-1mm}\right)$ & $
\epsilon^{2}
\hspace{-1mm}\left(\hspace{-1mm}\ba{ccc}
\epsilon^{2}&\epsilon &\epsilon^{2} \\ 1&\epsilon &  1 \\ 1&\epsilon^{3} & 1 \ea\hspace{-1mm}\right)$ & $\epsilon
\hspace{-1mm}\left(\hspace{-1mm}\ba{ccc}
1&\epsilon^{3} & 1 \\ \epsilon^{3}&\epsilon &\epsilon^{3} \\ 1&\epsilon^{3} & 1 \ea\hspace{-1mm}\right)$ & $\ba{c}
(2,6),(0,0),(0,1)\\(0,6),(1,1),(0,8)\\(1,0),(2,5),(1,0)\ea$
& $Z_{3}\times Z_{9}$\\[6.5mm]\hline
\multicolumn{6}{c}{}\\
\caption{\label{tab:models} A list of 22 valid flavor models for nearly
 tribimaximal lepton mixing. Shown are the explicit flavor charges
 under the flavor symmetry group $G_F$ and the resulting textures. Possible common overall suppression
 factors have been factored out of the textures.}
\end{longtable}
Some of the models in \Tab~\ref{tab:models} allow to extract very easily
new sum rules for the leptonic mixing angles using an expansion in $\epsilon$. For example, we find for model \#6 the
sum rules\\
\be
\theta_{12}=\frac{\pi}{4}-\frac{\epsilon}{\sqrt{2}}-\frac{\epsilon^2}{4},\quad\quad  \theta_{13}=(\frac{1}{\sqrt{2}}-\frac{1}{2})\epsilon^2,\quad\quad  \theta_{23}=\frac{\pi}{4}+\frac{\epsilon}{\sqrt{2}}+(\frac{1}{\sqrt{2}}-\frac{5}{4})\epsilon^2,\label{equ:model6}
\ee
while model \#8 exhibits the relations\\
\be
\theta_{12}=\frac{\pi}{4}-\frac{\epsilon}{\sqrt{2}}+(\frac{1}{\sqrt{2}}-\frac{9}{4})\epsilon^2,\quad\quad  \theta_{13}=\frac{\epsilon^2}{2},\quad\quad  \theta_{23}=\frac{\pi}{4}+\frac{\epsilon}{\sqrt{2}}+(\frac{1}{\sqrt{2}}-\frac{5}{4})\epsilon^2.\label{equ:model8}
\ee
For both the models \#6 and \#8, the solar angle satisfies the
well-known QLC relation $\theta_{12}\approx\frac{\pi}{4}-\theta_\text{C}/\sqrt{2}$. The
reactor angle, on the other hand, becomes for these two models small due to an apparent
suppression by a factor $\sim\theta_\text{C}^2$. Note that a similar suppression
of the reactor angle has been found for the model in
Ref.~\cite{Feruglio:2007hi}. The atmospheric angle follows in both
models \#6 and \#8 the new sum rule
$\theta_{23}\approx\frac{\pi}{4}+\theta_\text{C}/\sqrt{2}$, which
predicts a deviation from maximal mixing by an amount of approximately
$\theta_\text{C}/\sqrt{2}$. This prediction makes these models
testable in future neutrino oscillation experiments. For example, the
deviation from maximal mixing can be established at $3\sigma$ CL by the T2K or NO$\nu$A experiments \cite{Antusch:2004yx}. In addition, one
can measure the sign of the deviation from maximal mixing (the octant)
with a neutrino factory at $3\sigma$ CL for
$\text{sin}^22\theta_{13}\gtrsim 10^{-2.5}$ or at 90\% CL otherwise \cite{Gandhi:2006gu}.

The relations for $\theta_{13}$ in Eqs.~(\ref{equ:model6}) and
(\ref{equ:model8}), however, are not stable under variations of the Yukawa
couplings and must therefore be a result of exact cancellations between large
contributions from different sectors. To render these relations for $\theta_{13}$
natural, it might, thus, be necessary to extend these models by
non-Abelian discrete symmetries as demonstrated in Ref.~\cite{nonAbelian}.

Let us therefore have a look at model \#5, which has the sum rules
\be
\theta_{12}=\frac{\pi}{4}-\frac{\epsilon}{\sqrt{2}}-\frac{\epsilon^2}{4},\quad\quad  \theta_{13}=\mathcal{O}(\epsilon^3),\quad\quad  \theta_{23}=\frac{\pi}{4}+\frac{\epsilon}{\sqrt{2}}-\frac{3}{4}\epsilon^2.\label{equ:model5}
\ee
The sum rules for $\theta_{12}$ and $\theta_{23}$ are to leading order
just as in the two previous examples. The relation for $\theta_{13}$,
however, is different from that in Eqs.~(\ref{equ:model6}) and
(\ref{equ:model8}): Now, $\theta_{13}$ is very small due to a
suppression by a factor $\sim\theta_\text{C}^3$. Unlike for models \#6
and \#8, the relation for $\theta_{13}$ in \equ{model5} is stable under
variations of the Yukawa couplings. This means that by varying the
order one Yukawa couplings,
$\theta_{13}$ picks up only a small relative correction that is
suppressed by a factor $\sim\theta_\text{C}^3$.
\begin{figure}[th]
\begin{center}
\includegraphics[width=\textwidth]{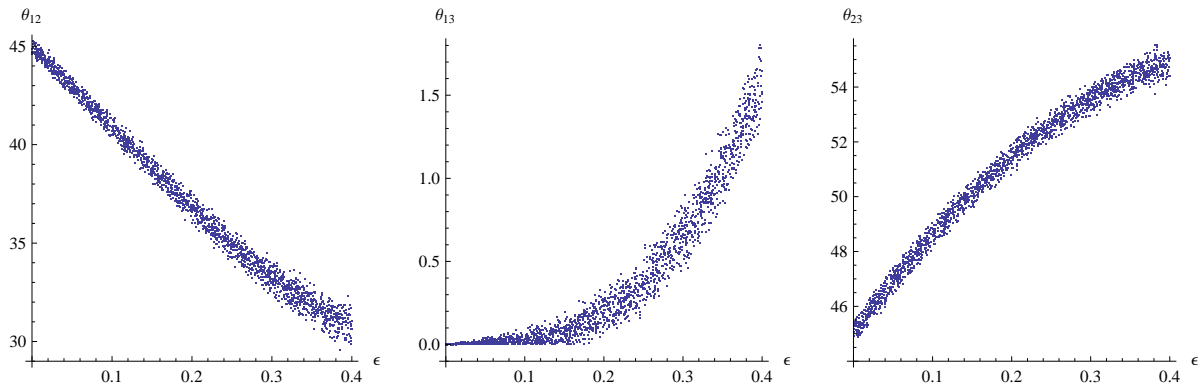}
\end{center}
\vspace*{-0.5cm}
\caption{\label{fig:stability} Solar (left),
  reactor (middle), and atmospheric (right) mixing angles (in degrees),
  for model \#5 in \Tab~\ref{tab:models} as a function of the
  expansion parameter $\epsilon$. The points correspond to a random
  variation of the order one Yukawa couplings in $Y_\ell',Y_D',$ and
  $Y_R'$, by about 1\%. The plot in the middle shows that the relation
  $\theta_{13}\simeq\mathcal{O}(\epsilon^3)$ in \equ{model5} is not
  due to an accidental cancellation between large mixing angles.}
\end{figure}
The stability of the relation
$\theta_{13}\simeq\mathcal{O}(\epsilon^3)$ under 1\% variations of the
order one Yukawa couplings is shown in \figu{stability}. While
$\theta_{12}$ and $\theta_{23}$ exhibit variations of the order
$(0.5-1)^\circ$ in the limit $\epsilon\rightarrow 0$, the variation of
$\theta_{13}$ remains $\ll 0.1^\circ$. Nevertheless, the sum rules for $\theta_{12}$ and
$\theta_{23}$ are in all three above examples in Eqs.~(\ref{equ:model6}),
(\ref{equ:model8}), and (\ref{equ:model5}), in this sense stable,
since the leading term in the respective expansions is $\pi/4$.

In analogy with the $Z_n$ groups, we have also performed a scan
of $U(1)$ symmetries. In order to see whether the cyclic character of
the discrete groups is a relevant feature in the construction
of valid models, we have compared single $Z_n$ groups $G_F=Z_{n_1}$ up to $|G_F|=40$
and product groups $G_F=Z_{n_1}\times Z_{n_2}$ up to $|G_F|=18$ with
corresponding Abelian groups, i.e., with $U(1)$ (for
$G_F=Z_{n_k}$) and $U(1)\times U(1)$ (for $G_F=Z_{n_1}\times Z_{n_2}$)
by letting (the absolute value of) the respective individual $U(1)$ charges vary
in the whole range
$0,1,\dots,n_k-1$. Thereby, we have found 129 models
for the cyclic groups and 24 models using $U(1)$ symmetries. All of
the 24 $U(1)$ models correspond to the group $G_F=Z_3\times Z_6$, for
which we have obtained in total 21 models. However, even though
the $U(1)$ models produce textures that are very similar to those of model \#6
in \Tab~\ref{tab:models} (with varying $M_\ell$ and varying first row
in $M_D$), $M_D$ and $M_R$ get in these examples highly suppressed by factors $\lesssim\epsilon^8$. This analysis already indicates that the cyclic character of
the flavor groups may be essential for facilitating the construction
of realistic flavor models.

\section{Summary and Conclusions}
\label{sec:summary}
In this paper, we have constructed several thousand
explicit flavor models for nearly
tribimaximal lepton mixing from products of $Z_n$ flavor
symmetries. In our models, small neutrino masses emerge only from the
canonical type-I seesaw mechanism. Upon flavor symmetry breaking, the
Froggatt-Nielsen mechanism produces hierarchical Yukawa coupling and
mass matrix textures of the leptons. These
textures are parameterized by powers of a single small symmetry breaking
parameter $\epsilon\simeq\theta_\text{C}\simeq 0.2$ that is of the order of the
Cabibbo angle $\theta_\text{C}$ and arises from integrating out heavy Froggatt-Nielsen
messenger fermions.

All flavor models realize the assumptions of EQLC and yield an
excellent fit to nearly tribimaximal neutrino mixing with a very small
reactor angle $\theta_{13}\approx 0$. Moreover, they lead to the hierarchical
charged lepton mass spectrum as well as to normal hierarchical
neutrino masses. In our analysis, we have
restricted ourselves to the most general CP-conserving case of real lepton mass
matrices. We have performed a
systematic scan of the group space for groups with up to four $Z_n$
factors and a maximum group order of 45. As a consequence, we have
found 6021 valid models that reproduce 2093 distinct texture sets.

A characteristic property of our flavor models is that large leptonic mixings
can come from the charged leptons and/or neutrinos. In the neutrino
sector, maximal mixings can arise in both the Dirac mass matrix or in the heavy right-handed
Majorana mass matrix. Generally, we can have maximal mixings between
any two generations in any lepton sector.

Among several explicit models, we have found a model that predicts a new relation $\theta_{13}=\mathcal{O}(\theta_\text{C}^3)$, for the
reactor angle. Moreover, we found models that all satisfy a new sum
rule $\theta_{23}\approx\frac{\pi}{4}+\theta_\text{C}/\sqrt{2}$ for
the atmospheric mixing angle, which makes these models testable in
future neutrino oscillation experiments such as the T2K and NO$\nu$A
experiments or at a neutrino factory.

We wish to point out that in this paper we have established a
connection between a model building top-down
approach using flavor symmetries and the phenomenological bottom-up approach of
Ref.~\cite{Plentinger:2007px}. While Ref.~\cite{Plentinger:2007px} deals with the
extraction of viable lepton mass textures that are in agreement with
observation, the current work successfully matches the
textures onto explicit flavor models, where the textures are
predicted from flavor symmetries and their breaking.

We believe that it would be interesting to study our sample of flavor
models with respect to further model building
aspects, anomaly cancellation, the inclusion of CP-violating phases,
as well as in view of lepton flavor violation or leptogenesis.

\section*{Appendix}
In this Appendix, we list the complete information on the mass and
mixing parameters of the matrices that are generated by the models in
\Tab~\ref{tab:models}. Here, ``\#'' labels in both tables the same
model. The data in \Tab~\ref{tab:modeldata} allows to fully
reconstruct the exact form of the mass matrices of the 22 models
following the notation of Sec.~\ref{sec:masses+mixings} (for further
detailed examples on such reconstructions, see also
Ref.~\cite{Plentinger:2007px}).

\begin{longtable}{@{\hspace{0mm}}c|@{\hspace{0mm}}c@{\hspace{0mm}}|@{\hspace{0mm}}c@{\hspace{0mm}}|@{\hspace{0mm}}c@{\hspace{0mm}}|@{\hspace{0mm}}c@{\hspace{0mm}}|@{\hspace{0mm}}c@{\hspace{0mm}}|@{\hspace{0mm}}c@{\hspace{0mm}}}
\# & $\ba{c}m_i^D/m_D\\m_i^R/M_{B-L}\ea$ & $\ba{c}(\theta_{12}^{\ell},\theta_{13}^{\ell},\theta_{23}^{\ell})\\(\delta^\ell,\alpha_1^\ell,\alpha_2^\ell)\ea$ & $\ba{c} (\theta_{12}^{\ell'},\theta_{13}^{\ell'},\theta_{23}^{\ell'})\\(\delta^{\ell'},\alpha_1^{\ell'},\alpha_2^{\ell'}) \ea$ &
$\ba{c}(\theta_{12}^{D},\theta_{13}^{D},\theta_{23}^{D}) \\ (\delta^D,\varphi_1^D,\varphi_2^D,\varphi_3^D) \ea$ &
$\ba{c} (\theta_{12}^{D'},\theta_{13}^{D'},\theta_{23}^{D'}) \\ (\delta^{D'},\alpha_1^{D'},\alpha_2^{D'}) \ea$
&
$\ba{c}(\theta_{12}^{R},\theta_{13}^{R},\theta_{23}^{R}) \\ (\delta^R,\varphi_1^R,\varphi_2^R,\varphi_3^R) \ea$
\\[2mm]\hline\hline&&&&&&\\*[-4mm]
\endfirsthead

1&
$\begin{array}{c}
( \epsilon  , 1 , \epsilon  ) \\
( \epsilon  , 1 , 1 )
\end{array}$
&
 $\begin{array}{c}
  ( \epsilon ^2 , \epsilon ^2 , \epsilon ^2 ) \\
  ( \pi  , \pi  , \pi  )
 \end{array}$
&
 $\begin{array}{c}
( \frac{\pi }{4} , \epsilon ^2 , 0 ) \\
  ( 0 , 0 , 0 )
 \end{array}$
&$\ba{c}
  ( 0 , \epsilon ^2 , \frac{\pi }{4} ) \\
  ( 0 , 0 , 0 , 0 )
\ea$
&$\ba{c}
  ( \epsilon  , \epsilon  , \epsilon ^2 ) \\
  ( 0 , 0 , 0 )
\ea$
&$\ba{c}
  ( \epsilon ^2 , \frac{\pi }{4} , \epsilon ^2 ) \\
  ( \pi  , 0 , \pi  , 0 )
\ea$
\\[2mm]\hline&&&&&&\\*[-4mm]

2&
$\begin{array}{c}
( \epsilon  , 1 , \epsilon  ) \\
( \epsilon  , 1 , 1 )
\end{array}$
&
 $\begin{array}{c}
  ( \epsilon ^2 , \epsilon ^2 , \frac{\pi }{4} ) \\
      ( 0 , 0 , 0 )
 \end{array}$
&
 $\begin{array}{c}
( \epsilon  , 0 , \epsilon ^2 ) \\
  ( 0 , 0 , 0 )
 \end{array}$&$
\ba{c}
  ( \epsilon ^2 , \frac{\pi }{4} , \epsilon ^2 )
\\
  ( 0 , 0 , 0 , 0 )
\ea$
&$\ba{c}
  ( \epsilon  , \epsilon ^2 , 0 )
\\
  ( 0 , \pi  , \pi  )
\ea$
&$\ba{c}
  ( \epsilon  , 0 , \frac{\pi }{4} )
\\
  ( 0 , 0 , 0 , 0 )
\ea$
\\[2mm]\hline&&&&&&\\*[-4mm]

3&$
\begin{array}{c}
( \epsilon  , 1 , \epsilon  ) \\
( \epsilon  , 1 , 1 )
\end{array}
$&$
 \begin{array}{c}
  ( \epsilon  , 0 , \epsilon ^2 ) \\
   ( 0 , 0 , 0 )
 \end{array}
$&$
  \begin{array}{c}
  ( \epsilon  , \epsilon  , \epsilon ^2 ) \\
   ( 0 , 0 , 0 )
  \end{array}$
&$\ba{c}
  ( \epsilon  , \frac{\pi }{4} , \frac{\pi }{4} )
\\
  ( \pi  , 0 , 0 , \pi  )
\ea$
&$\ba{c}
  ( \epsilon  , \epsilon  , \frac{\pi }{4} )
\\
   ( 0 , \pi  , \pi  )
\ea$
&$\ba{c}
  ( \epsilon  , \epsilon  , \frac{\pi }{4} )
\\
   ( 0 , 0 , 0 , \pi  )
\ea$
\\[2mm]\hline&&&&&&\\*[-4mm]

4&$
\begin{array}{c}
( \epsilon  , \epsilon  , 1 ) \\
( \epsilon  , 1 , 1 )
\end{array}
$&$
 \begin{array}{c}
  ( \epsilon  , \epsilon  , \epsilon ^2 ) \\
   ( 0 , \pi  , \pi  )
 \end{array}
$&$
  \begin{array}{c}
  ( \frac{\pi }{4} , \epsilon ^2 , 0 ) \\
   ( 0 , 0 , 0 )
  \end{array}$
&$\ba{c}
  ( \frac{\pi }{4} , \epsilon ^2 , \frac{\pi }{4} )
    \\
   ( 0 , 0 , \pi  , 0 )
\ea$
&$\ba{c}
  ( 0 , \epsilon ^2 , \frac{\pi }{4} ) \\
   ( 0 , 0 , 0 )
\ea$
&$\ba{c}
  ( 0 , \epsilon  , \epsilon ^2 )
\\
   ( 0 , 0 , \pi  , 0 )
\ea$
\\[2mm]\hline&&&&&&\\*[-4mm]
5&$
\begin{array}{c}
( \epsilon  , 1 , \epsilon  ) \\
( \epsilon  , 1 , 1 )
\end{array}
$&$
 \begin{array}{c}
  ( 0 , \epsilon  , \epsilon ^2 ) \\     ( 0 , \pi  , 0 )
 \end{array}
$&$
  \begin{array}{c}
  ( \epsilon  , 0 , \epsilon  ) \\
   ( 0 , 0 , 0 )
  \end{array}$
&$\ba{c}
  ( \epsilon  , \frac{\pi }{4} , \frac{\pi }{4} )
\\
   ( \pi  , 0 , 0 , \pi  )
\ea$
&$\ba{c}
  ( \epsilon ^2 , \frac{\pi }{4} , \epsilon ^2 )
\\
   ( 0 , \pi  , \pi  )
\ea$
&$\ba{c}
  ( 0 , \frac{\pi }{4} , 0 )
\\
   ( 0 , 0 , 0 , \pi  )
\ea$
\\[2mm]\hline&&&&&&\\*[-4mm]

6&$
\begin{array}{c}
( \epsilon ^2 , 1 , \epsilon  ) \\
( \epsilon ^2 , \epsilon  , 1 )
\end{array}
$&$
 \begin{array}{c}
  ( 0 , \epsilon  , \epsilon ^2 ) \\     ( 0 , \pi  , 0 )
 \end{array}
$&$
  \begin{array}{c}
  ( \epsilon  , 0 , \epsilon  ) \\
   ( 0 , 0 , 0 )
 \end{array}$
&$\ba{c}
  ( \epsilon  , \frac{\pi }{4} , \frac{\pi }{4} )
    \\
   ( \pi  , 0 , 0 , \pi  )
\ea$
&$\ba{c}
  ( \epsilon  , \epsilon  , \frac{\pi }{4} ) \\
   ( \pi  , 0 , 0 )
\ea$
&$\ba{c}
  ( \epsilon  , \epsilon  , \frac{\pi }{4} )
\\
   ( \pi  , 0 , \pi  , 0 )
\ea$
\\[2mm]\hline&&&&&&\\*[-4mm]

7&$
\begin{array}{c}
( \epsilon ^2 , \epsilon  , 1 ) \\
( \epsilon ^2 , \epsilon  , 1 )
\end{array}
$&$
 \begin{array}{c}
  ( 0 , \epsilon ^2 , \frac{\pi }{4} ) \\    ( 0 , \pi  , \pi  )
 \end{array}
$&$
 \begin{array}{c}
  ( \epsilon  , 0 , \epsilon ^2 ) \\
  ( 0 , 0 , \pi  )
 \end{array}$
&$\ba{c}
  ( \frac{\pi }{4} , \epsilon ^2 , \epsilon  ) \\
  ( 0 , 0 , 0 , \pi  )
\ea$
&$\ba{c}
  ( 0 , \frac{\pi }{4} , \epsilon  ) \\
  ( 0 , 0 , \pi  )
\ea$
&$\ba{c}
  ( \epsilon ^2 , \frac{\pi }{4} , \epsilon ^2 )
\\
  ( \pi  , 0 , \pi  , \pi  )
\ea$
\\[2mm]\hline&&&&&&\\*[-4mm]

8&$
\begin{array}{c}
( \epsilon ^2 , 1 , \epsilon  ) \\
( \epsilon ^2 , \epsilon  , 1 )
\end{array}
$&$
 \begin{array}{c}
  ( \epsilon ^2 , \epsilon  , \epsilon ^2 ) \\     ( 0 , \pi  , 0 )
 \end{array}
$&$
  \begin{array}{c}
  ( \epsilon  , \epsilon ^2 , 0 ) \\
   ( 0 , 0 , 0 )
  \end{array}
$
&$\ba{c}
  ( \epsilon  , \frac{\pi }{4} , \frac{\pi }{4} )
    \\
   ( \pi  , 0 , 0 , \pi  )
\ea$
&$\ba{c}
  ( 0 , \epsilon ^2 , \frac{\pi }{4} ) \\
   ( 0 , 0 , \pi  )
\ea$
&$\ba{c}
  ( \epsilon ^2 , \epsilon ^2 , \frac{\pi }{4} )
\\
   ( \pi  , 0 , \pi  , 0 )
\ea$\\[2mm]\hline&&&&&&\\*[-4mm]

9&$

\begin{array}{c}
( \epsilon  , 1 , \epsilon  ) \\
( \epsilon  , 1 , 1 )
\end{array}
$&$
 \begin{array}{c}
  ( \epsilon  , 0 , \epsilon ^2 ) \\     ( 0 , 0 , 0 )
 \end{array}
$&$
  \begin{array}{c}
  ( \epsilon  , \epsilon  , 0 ) \\
   ( 0 , 0 , 0 )
  \end{array}
$
&$\ba{c}
  ( \epsilon  , \frac{\pi }{4} , \frac{\pi }{4} )
    \\
   ( \pi  , 0 , 0 , \pi  )
\ea$
&$\ba{c}
  ( \frac{\pi }{4} , \epsilon  , \epsilon ^2 ) \\
   ( \pi  , 0 , \pi  )
\ea$
&$\ba{c}
  ( \frac{\pi }{4} , \epsilon  , 0 )
\\
   ( 0 , 0 , 0 , \pi  )
\ea$\\[2mm]\hline&&&&&&\\*[-4mm]

10&$

\begin{array}{c}
( \epsilon ^2 , \epsilon  , 1 ) \\
( \epsilon ^2 , \epsilon  , 1 )
\end{array}
$&$
 \begin{array}{c}
  ( \epsilon ^2 , \epsilon ^2 , \epsilon ^2 ) \\     ( \pi  , \pi  , \pi  )
 \end{array}
$&$
  \begin{array}{c}
  ( \frac{\pi }{4} , \epsilon ^2 , 0 ) \\
   ( 0 , 0 , 0 )
  \end{array}
$
&$\ba{c}
  ( \frac{\pi }{4} , \epsilon ^2 , \frac{\pi }{4} )
    \\
   ( 0 , 0 , 0 , \pi  )
\ea$
&$\ba{c}
  ( \epsilon ^2 , 0 , \epsilon ^2 ) \\
   ( 0 , 0 , 0 )
\ea$
&$\ba{c}
  ( \epsilon  , \epsilon ^2 , 0 )
\\
   ( 0 , 0 , 0 , 0 )
\ea$\\[2mm]\hline&&&&&&\\*[-4mm]

11&$

\begin{array}{c}
( \epsilon  , 1 , \epsilon  ) \\
( \epsilon  , 1 , 1 )
\end{array}
$&$
 \begin{array}{c}
  ( \epsilon  , \epsilon ^2 , \epsilon ^2 ) \\     ( \pi  , 0 , 0 )
 \end{array}
$&$
  \begin{array}{c}
  ( \frac{\pi }{4} , 0 , \epsilon ^2 ) \\
   ( 0 , 0 , 0 )
  \end{array}
$
&$\ba{c}
  ( \epsilon  , \frac{\pi }{4} , \frac{\pi }{4} )
    \\
   ( \pi  , 0 , 0 , \pi  )
\ea$
&$\ba{c}
  ( \epsilon  , \epsilon ^2 , \epsilon  ) \\
   ( \pi  , 0 , 0 )
\ea$
&$\ba{c}
  ( \epsilon  , \epsilon ^2 , \epsilon  )
\\
   ( \pi  , 0 , 0 , 0 )
\ea$\\[2mm]\hline&&&&&&\\*[-4mm]
12&$

\begin{array}{c}
( \epsilon  , 1 , \epsilon  ) \\
( \epsilon  , 1 , 1 )
\end{array}
$&$
 \begin{array}{c}
  ( 0 , \epsilon  , \epsilon ^2 ) \\
   ( 0 , \pi  , 0 )
 \end{array}
$&$
  \begin{array}{c}
  ( \epsilon  , 0 , 0 ) \\
   ( 0 , 0 , 0 )
  \end{array}
$&$\ba{c}
  ( \epsilon  , \frac{\pi }{4} , \frac{\pi }{4} )
\\
   ( \pi  , 0 , 0 , \pi  )
\ea$
&$\ba{c}
  ( \epsilon ^2 , \frac{\pi }{4} , \epsilon  )
\\
   ( 0 , \pi  , 0 )
\ea$
&$\ba{c}
  ( \epsilon  , \frac{\pi }{4} , \epsilon ^2 )
\\
  ( \pi  , 0 , \pi  , 0 )
\ea$
\\[2mm]\hline&&&&&&\\*[-4mm]
13&$

\begin{array}{c}
( \epsilon ^2 , \epsilon  , 1 ) \\
( \epsilon ^2 , \epsilon  , 1 )
\end{array}
$&$
 \begin{array}{c}
  ( \epsilon ^2 , \epsilon ^2 , \epsilon ^2 ) \\
   ( \pi  , 0 , \pi  )

 \end{array}
$&$
  \begin{array}{c}
  ( \epsilon  , 0 , 0 ) \\
   ( 0 , 0 , 0 )
  \end{array}
$
&$\ba{c}
  ( \frac{\pi }{4} , \epsilon ^2 , \frac{\pi }{4} )
\\
   ( \pi  , 0 , \pi  , 0 )
\ea$
&$\ba{c}
  ( \epsilon  , \epsilon ^2 , 0 )
\\
   ( 0 , \pi  , 0 )
\ea$
&$\ba{c}
  ( 0 , \epsilon ^2 , \epsilon ^2 )
\\
   ( 0 , 0 , \pi  , \pi  )
\ea$\\[2mm]\hline&&&&&&\\*[-4mm]
14&$

\begin{array}{c}
( \epsilon ^2 , \epsilon  , 1 ) \\
( \epsilon ^2 , \epsilon  , 1 )
\end{array}
$&$
 \begin{array}{c}
  ( \epsilon ^2 , \epsilon ^2 , \frac{\pi }{4} ) \\
  ( \pi  , \pi  , \pi  )   \end{array}
$&$
 \begin{array}{c}
  ( \epsilon  , 0 , \epsilon ^2 ) \\
  ( 0 , 0 , \pi  )   \end{array}
$
&$\ba{c}
  ( \frac{\pi }{4} , \epsilon ^2 , 0 )
\\
  ( 0 , 0 , 0 , 0 )
\ea$
&$\ba{c}
  ( \epsilon ^2 , \frac{\pi }{4} , 0 )
\\
  ( 0 , 0 , \pi  )
\ea$
&$\ba{c}
  ( \epsilon ^2 , \frac{\pi }{4} , \epsilon  )
\\
  ( 0 , 0 , \pi  , 0 )
\ea$\\[2mm]\hline&&&&&&\\*[-4mm]
15&$

\begin{array}{c}
( \epsilon ^2 , \epsilon  , 1 ) \\
( \epsilon ^2 , \epsilon  , 1 )
\end{array}
$&$
 \begin{array}{c}
  ( \epsilon ^2 , \epsilon ^2 , \frac{\pi }{4} ) \\    ( \pi  , \pi  , \pi  )
 \end{array}
$&$
 \begin{array}{c}
  ( \epsilon  , 0 , \epsilon ^2 )\\
  ( 0 , 0 , 0 )
 \end{array}
$
&$\ba{c}
  ( \frac{\pi }{4} , 0 , \epsilon ^2 )
\\
  ( 0 , 0 , 0 , \pi  )
\ea$
&$\ba{c}
  ( \epsilon ^2 , \epsilon  , \epsilon ^2 )
\\
  ( 0 , 0 , 0 )
\ea$
&$\ba{c}
  ( \epsilon  , \epsilon  , \epsilon ^2 )
\\
  ( \pi  , 0 , 0 , \pi  )
\ea$\\[2mm]\hline&&&&&&\\*[-4mm]

16&$

\begin{array}{c}
( \epsilon  , 1 , \epsilon ) \\
( \epsilon ,1  , 1 )
\end{array}
$&$
 \begin{array}{c}
  ( \epsilon  , \epsilon ^2 , \frac{\pi }{4} ) \\    ( \pi  , 0 , 0 )
 \end{array}
$&$
 \begin{array}{c}
  ( \epsilon  , 0 , \epsilon ^2 ) \\
  ( 0 , 0 , 0 )
 \end{array}
$
&$\ba{c}
  ( \epsilon  , \frac{\pi }{4} , \epsilon ^2 ) \\
  ( \pi  , 0 , 0 , \pi  )
\ea$
&$\ba{c}
  ( \frac{\pi }{4} , 0,0 ) \\
  ( 0 , 0 , \pi )
\ea$
&$\ba{c}
  ( \frac{\pi }{4} , \epsilon^2 , \epsilon^2 )
\\
  ( 0 , 0 , 0  , 0 )
\ea$\\[2mm]\hline&&&&&&\\*[-4mm]
17&$

\begin{array}{c}
( \epsilon ^2 , \epsilon  , 1 ) \\
( \epsilon ^2 , \epsilon  , 1 )
\end{array}
$&$
 \begin{array}{c}
  ( \epsilon ^2 , \epsilon ^2 , \frac{\pi }{4} ) \\    ( \pi  , \pi  , \pi  )
 \end{array}
$&$
 \begin{array}{c}
  ( \epsilon  , 0 , \epsilon ^2 ) \\
  ( 0 , 0 , 0 )
 \end{array}
$
&$\ba{c}
  ( \frac{\pi }{4} , 0 , \epsilon ^2 ) \\
  ( 0 , 0 , 0 , \pi  )
\ea$
&$\ba{c}
  ( \epsilon ^2 , 0 , \epsilon ^2 ) \\
  ( 0 , 0 , 0 )
\ea$
&$\ba{c}
  ( \epsilon  , 0 , \epsilon ^2 )
\\
  ( 0 , 0 , 0 , \pi  )
\ea$\\[2mm]\hline&&&&&&\\*[-4mm]
18&$

\begin{array}{c}
( \epsilon ^2 , \epsilon  , 1 ) \\
( \epsilon ^2 , \epsilon  , 1 )
\end{array}
$&$
 \begin{array}{c}
  ( \epsilon ^2 , \epsilon ^2 , \epsilon ^2 ) \\     ( \pi  , 0 , \pi  )
 \end{array}
$&$
  \begin{array}{c}
  ( \epsilon  , 0 , \epsilon ^2 ) \\
   ( 0 , 0 , 0 )
  \end{array}
$
&$\ba{c}
  ( \frac{\pi }{4} , \epsilon ^2 , \frac{\pi }{4} )
    \\
   ( \pi  , 0 , \pi  , 0 )
\ea$
&$\ba{c}
  ( \epsilon  , \epsilon ^2 , \epsilon ^2 ) \\
   ( \pi  , 0 , \pi  )
\ea$
&$\ba{c}
  ( 0 , \epsilon ^2 , 0 )
\\
   ( 0 , 0 , 0 , 0 )
\ea$\\[2mm]\hline&&&&&&\\*[-4mm]
19&$

\begin{array}{c}
( \epsilon  , 1 , \epsilon  ) \\
( \epsilon  , 1 , 1 )
\end{array}
$&$
 \begin{array}{c}
  ( \epsilon ^2 , \epsilon ^2 , \epsilon ^2 ) \\     ( \pi  , \pi  , \pi  )
 \end{array}
$&$
  \begin{array}{c}
  ( \frac{\pi }{4} , 0 , \epsilon ^2 ) \\
   ( 0 , 0 , 0 )
  \end{array}
$
&$\ba{c}
  ( \epsilon ^2 , \frac{\pi }{4} , \frac{\pi }{4} )
    \\
   ( \pi  , 0 , 0 , 0 )
\ea$
&$\ba{c}
  ( 0 , \epsilon  , \epsilon  ) \\
   ( 0 , 0 , \pi  )
\ea$
&$\ba{c}
  ( \epsilon ^2 , \epsilon ^2 , \frac{\pi }{4} )
\\
   ( 0 , 0 , 0 , 0 )
\ea$\\[2mm]\hline&&&&&&\\*[-4mm]
20&$

\begin{array}{c}
( \epsilon ^2 , \epsilon  , 1 ) \\
( \epsilon ^2 , \epsilon  , 1 )
\end{array}
$&$
 \begin{array}{c}
  ( \epsilon ^2 , \epsilon ^2 , \epsilon ^2 ) \\     ( 0 , \pi  , \pi  )
 \end{array}
$&$
  \begin{array}{c}
  ( \frac{\pi }{4} , 0 , 0 ) \\
   ( 0 , 0 , 0 )
  \end{array}
$
&$\ba{c}
  ( \frac{\pi }{4} , \epsilon ^2 , \frac{\pi }{4} )
    \\
   ( \pi  , 0 , 0 , \pi  )
\ea$
&$\ba{c}
  ( \epsilon ^2 , \epsilon ^2 , \epsilon ^2 ) \\
   ( 0 , 0 , 0 )
\ea$
&$\ba{c}
  ( \epsilon  , \epsilon ^2 , \epsilon ^2 )
\\
   ( \pi  , 0 , 0 , \pi  )
\ea$\\[2mm]\hline&&&&&&\\*[-4mm]
21&$

\begin{array}{c}
( \epsilon  , 1 , \epsilon  ) \\
( \epsilon  , 1 , 1 )
\end{array}
$&$
 \begin{array}{c}
  ( \epsilon ^2 , \epsilon ^2 , \epsilon ^2 ) \\    ( \pi  , \pi  , \pi  )
 \end{array}
$&$
 \begin{array}{c}
  ( \frac{\pi }{4} , 0 , 0 ) \\
  ( 0 , 0 , 0 )
 \end{array}
$
&$\ba{c}
  ( 0 , \epsilon ^2 , \frac{\pi }{4} ) \\
  ( 0 , 0 , 0 , 0 )
\ea$
&$\ba{c}
  ( \epsilon  , \frac{\pi }{4} , \epsilon  ) \\
  ( 0 , 0 , 0 )
\ea$
&$\ba{c}
  ( \epsilon ^2 , \epsilon  , \epsilon ^2 )
\\
  ( \pi  , 0 , \pi  , 0 )
\ea$\\[2mm]\hline&&&&&&\\*[-4mm]
22&$

\begin{array}{c}
( \epsilon ^2 , \epsilon  , 1 ) \\
( \epsilon ^2 , \epsilon  , 1 )
\end{array}
$&$
 \begin{array}{c}
  ( \epsilon ^2 , \epsilon ^2 , 0 ) \\     ( 0 , 0 , 0 )
 \end{array}
$&$
  \begin{array}{c}
  ( \frac{\pi }{4} , 0 , \epsilon  ) \\
   ( 0 , 0 , 0 )
  \end{array}
$&$\ba{c}
  ( \frac{\pi }{4} , \epsilon ^2 , \frac{\pi }{4} )
    \\
   ( \pi  , 0 , 0 , 0 )
\ea$
&$\ba{c}
  ( \epsilon ^2 , \frac{\pi }{4} , \epsilon  ) \\
   ( 0 , \pi  , 0 )
\ea$
&$\ba{c}
  ( \epsilon ^2 , \frac{\pi }{4} , 0 )
\\
   ( 0 , 0 , \pi  , 0 )
\ea$
\\[2mm]\hline
\multicolumn{7}{c}{}\\
\caption{\label{tab:modeldata}Supplementary information for the
 reconstruction of the mass matrices and Yukawa couplings of the
 models in \Tab~\ref{tab:models}. Note that we have made use of the
 freedom to set
 $\varphi_i^\ell=\varphi_i^{\ell'}=\varphi_i^{D'}=\alpha_j^D=0$ for
 $i=1,2,3$ and $j=1,2$.}
\end{longtable}

\section*{Acknowledgments}

The research of F.P. is supported by Research Training Group 1147 \textit{Theoretical Astrophysics and Particle Physics} of Deutsche Forschungsgemeinschaft.
G.S. is supported by the Federal Ministry of Education and Research (BMBF) under contract number 05HT1WWA2.
W.W. would like to acknowledge support from the Emmy Noether program of Deutsche Forschungsgemeinschaft.


\begin{thebibliography}{00}

\bibitem{Fukuda:2002pe}
  S.~Fukuda {\it et al.}  [Super-Kamiokande Collaboration],
  Phys.\ Lett.\  B {\bf 539}, 179 (2002)
  [arXiv:hep-ex/0205075]; Q.~R.~Ahmad {\it et al.}  [SNO Collaboration],
  Phys.\ Rev.\ Lett.\  {\bf 89}, 011302 (2002)
  [arXiv:nucl-ex/0204009].

\bibitem{Fukuda:1998mi}
  Y.~Fukuda {\it et al.}  [Super-Kamiokande Collaboration],
  Phys.\ Rev.\ Lett.\  {\bf 81}, 1562 (1998)
  [arXiv:hep-ex/9807003].

\bibitem{Araki:2004mb}
  T.~Araki {\it et al.}  [KamLAND Collaboration],
  Phys.\ Rev.\ Lett.\  {\bf 94}, 081801 (2005)
  [arXiv:hep-ex/0406035];
  M.~Apollonio {\it et al.}  [CHOOZ Collaboration],
  Eur.\ Phys.\ J.\  C {\bf 27}, 331 (2003)
  [arXiv:hep-ex/0301017].

\bibitem{Aliu:2004sq}
  E.~Aliu {\it et al.}  [K2K Collaboration],
  Phys.\ Rev.\ Lett.\  {\bf 94}, 081802 (2005)
  [arXiv:hep-ex/0411038].

\bibitem{SU5}
H.~Georgi and S.~L.~Glashow, Phys. Rev. Lett. {\bf 32}, 438
  (1974); H.~Georgi, in {\it Proceedings of Coral Gables 1975, Theories and
  Experiments in High Energy Physics}, New York, 1975.

\bibitem{Pati:1974yy}
  J.~C.~Pati and A.~Salam,
  Phys.\ Rev.\  D {\bf 10}, 275 (1974)
  [Erratum-ibid.\  D {\bf 11}, 703 (1975)].

\bibitem{typeIseesaw}
  P.~Minkowski, Phys.\ Lett.\  B {\bf 67}, 421 (1977); T.~Yanagida, in {\it
  Proceedings of the Workshop on the Unified Theory and Baryon Number in the
  Universe}, KEK, Tsukuba, 1979; M.~Gell-Mann, P.~Ramond and R.~Slansky, in
  {\it Proceedings of the Workshop on Supergravity}, Stony Brook, New York,
  1979; S.~L.~Glashow, in {\it Proceedings of the 1979 Cargese Summer
  Institute on Quarks and Leptons}, New York, 1980.

\bibitem{typeIIseesaw}
M.~Magg and C.~Wetterich, Phys.\ Lett.\ B {\bf 94}, 61 (1980);
  R.~N.~Mohapatra and G.~Senjanovi\'c, Phys.\ Rev.\ Lett.\ {\bf 44},
  912 (1980); Phys.\ Rev.\ D {\bf 23}, 165 (1981); J.~Schechter and J.~W.~F.~Valle,
  Phys.\ Rev.\ D {\bf 22}, 2227 (1980); G.~Lazarides, Q.~Shafi and
  C.~Wetterich, Nucl.\ Phys.\ B {\bf 181}, 287 (1981).

\bibitem{GUTscale}
H.~Georgi and H.~Quinn, Phys.\ Rev.\ Lett.\ {\bf 33}, 451
  (1974); S.~Dimopoulos, S.~Raby and F.~Wilczek, Phys.\ Rev.\ D {\bf 24}, 1681
  (1981); S.~Dimopoulos and H.~Georgi, Nucl.\ Phys.\ B {\bf 193}, 150 (1981).

\bibitem{Schwetz:2006dh}
  T.~Schwetz,
  Phys.\ Scripta {\bf T127}, 1 (2006)
  [arXiv:hep-ph/0606060].

\bibitem{PMNS}
B.~Pontecorvo, Sov.\ Phys.\ JETP {\bf 6}, 429 (1957); Z.~Maki,
  M.~Nakagawa and S.~Sakata, Prog.\ Theor.\ Phys.\ {\bf 28}, 870 (1962).

\bibitem{Harrison:1999cf}
  P.~F.~Harrison, D.~H.~Perkins and W.~G.~Scott,
  Phys.\ Lett.\  B {\bf 458}, 79 (1999)
  [arXiv:hep-ph/9904297];
  P.~F.~Harrison, D.~H.~Perkins and W.~G.~Scott,
  Phys.\ Lett.\  B {\bf 530}, 167 (2002)
  [arXiv:hep-ph/0202074].

\bibitem{Plentinger:2005kx}
  F.~Plentinger and W.~Rodejohann,
  Phys.\ Lett.\  B {\bf 625}, 264 (2005)
  [arXiv:hep-ph/0507143].

\bibitem{Majumdar:2006px}
  D.~Majumdar and A.~Ghosal,
  Phys.\ Rev.\  D {\bf 75}, 113004 (2007)
  [arXiv:hep-ph/0608334];
  A.~H.~Chan, H.~Fritzsch, S.~Luo and Z.~z.~Xing,
  Phys.\ Rev.\  D {\bf 76}, 073009 (2007)
  [arXiv:0704.3153 [hep-ph]];
  S.~F.~King,
  Phys.\ Lett.\  B {\bf 659}, 244 (2008)
  [arXiv:0710.0530 [hep-ph]].

\bibitem{Xing:2002sw}
  Z.~z.~Xing,
  Phys.\ Lett.\  B {\bf 533}, 85 (2002)
  [arXiv:hep-ph/0204049];
  Z.~z.~Xing, H.~Zhang and S.~Zhou,
  Phys.\ Lett.\  B {\bf 641}, 189 (2006)
  [arXiv:hep-ph/0607091];
  E.~Ma, arXiv:0709.0507 [hep-ph];
  A.~Mondragon, M.~Mondragon and E.~Peinado,
  arXiv:0712.2488 [hep-ph].

\bibitem{A4}
  E.~Ma and G.~Rajasekaran,
  Phys.\ Rev.\  D {\bf 64}, 113012 (2001)
  [arXiv:hep-ph/0106291];
  K.~S.~Babu, E.~Ma and J.~W.~F.~Valle,
  Phys.\ Lett.\  B {\bf 552}, 207 (2003)
  [arXiv:hep-ph/0206292]; M.~Hirsch, J.~C.~Romao, S.~Skadhauge, J.~W.~F.~Valle and A.~Villanova del Moral,
  Phys.\ Rev.\  D {\bf 69}, 093006 (2004)
  [arXiv:hep-ph/0312265].

\bibitem{2A4}
  P.~H.~Frampton and T.~W.~Kephart,
  Int.\ J.\ Mod.\ Phys.\  A {\bf 10}, 4689 (1995)
  [arXiv:hep-ph/9409330];
  A.~Aranda, C.~D.~Carone and R.~F.~Lebed,
  Phys.\ Rev.\  D {\bf 62}, 016009 (2000)
  [arXiv:hep-ph/0002044];
  P.~D.~Carr and P.~H.~Frampton,
  arXiv:hep-ph/0701034; A.~Aranda,
  Phys.\ Rev.\  D {\bf 76}, 111301 (2007)
  [arXiv:0707.3661 [hep-ph]].

\bibitem{Altarelli:2007cd}
G.~Altarelli, arXiv:0705.0860 [hep-ph].

\bibitem{CKM}
N.~Cabibbo, Phys.\ Rev.\ Lett.\ {\bf 10}, 531 (1963); M.~Kobayashi and T.~Maskawa, Prog.\ Theor.\ Phys.\ {\bf 49}, 652 (1973).

\bibitem{discretequarks}
E.~Ma, Mod.\ Phys.\ Lett.\ A {\bf 17}, 627 (2002)
[arXiv:hep-ph/0203238]; G.~Altarelli and F.~Feruglio, Nucl.\ Phys.\ B
{\bf 741}, 215 (2006) [arXiv:hep-ph/0512103]; C.~Hagedorn, M.~Lindner
and F.~Plentinger, Phys.\ Rev.\  D {\bf 74}, 025007 (2006)
[arXiv:hep-ph/0604265]; S.~F.~King and M.~Malinsky, JHEP {\bf 11}, 071 (2006) [arXiv:hep-ph/0608021]; Phys.\
Lett.\ B {\bf 645}, 351 (2007) [arXiv:hep-ph/0610250]; F.~Feruglio,
C.~Hagedorn, Y.~Lin and L.~Merlo, Nucl.\ Phys.\ B {\bf 775}, 120
(2007) [arXiv:hep-ph/0702194]; C.~Luhn, S.~Nasri and P.~Ramond,
Phys.\ Lett.\  B {\bf 652}, 27 (2007) [[arXiv:0706.2341 [hep-ph]].

\bibitem{discreteGUTs}
E.~Ma, H.~Sawanaka and M.~Tanimoto, Phys.\ Lett.\ B {\bf 641},
  301 (2006) [arXiv:hep-ph/0606103]; I.~de Medeiros Varzielas, S.~F.~King and
  G.~G.~Ross, Phys.\ Lett.\ B {\bf 648}, 201 (2007)
  [arXiv:hep-ph/0607045]; E.~Ma, Mod.\ Phys.\ Lett.\ A {\bf 21}, 2931
  (2006) [arXiv:hep-ph/0607190]; S.~Morisi, M.~Picariello and
  E.~Torrente-Lujan, Phys.\ Rev.\ D {\bf 75}, 075015 (2007) [arXiv:hep-ph/0702034]; M.-C.~Chen and K.~T.~Mahanthappa, [arXiv:0705.0714
  [hep-ph]]; W.~Grimus and H.~Kuhbock, arXiv:0710.1585 [hep-ph]; G.~Altarelli, F.~Feruglio and C.~Hagedorn, arXiv:0802.0090 [hep-ph].

\bibitem{qlc}
  A.~Y.~Smirnov, arXiv:hep-ph/0402264;
  M.~Raidal,
  Phys.\ Rev.\ Lett.\  {\bf 93}, 161801 (2004)
  [arXiv:hep-ph/0404046];
  H.~Minakata and A.~Y.~Smirnov, Phys.\ Rev.\  D {\bf 70}, 073009 (2004)
  [arXiv:hep-ph/0405088].

\bibitem{qlcbimax}
M.~Jezabek and Y.~Sumino, Phys.\ Lett.\ B {\bf 457}, 139
  (1999) [arXiv: hep-ph/9904382]; C.~Giunti and M.~Tanimoto, Phys.\
Rev.\ D {\bf 66}, 113006 (2002) [arXiv:hep-ph/0209169];
P.~H.~Frampton, S.~T.~Petcov and W.~Rodejohann, Nucl.\ Phys.\ B {\bf
  687}, 31 (2004) [arXiv:hep-ph/0401206].

\bibitem{qlcsumrules}
T.~Ohlsson, Phys.\ Lett.\ B {\bf 622}, 159 (2005)
[arXiv:hep-ph/0506094]; S.~Antusch and S.~F.~King, Phys.\ Lett.\ B
{\bf 631}, 42 (2005) [arXiv:hep-ph/0508044].

\bibitem{qlcpheno}
K.~Cheung, S.~K.~Kang, C.~S.~Kim and J.~Lee, Phys.\ Rev.\ D {\bf 72},
036003 (2005) [arXiv:hep-ph/0503122]; K.~A.~Hochmuth and W.~Rodejohann,
  Phys.\ Rev.\ D {\bf 75}, 073001 (2007) [arXiv:hep-ph/0607103].

\bibitem{qlcCabibbo}
W.~Rodejohann, Phys.\ Rev.\ D {\tt 69}, 033005 (2004)
[arXiv:hep-ph/0309249]; N.~Li and B.-Q.~Ma, Phys.\ Rev.\ D {\bf 71},
097301 (2005) [arXiv:hep-ph/0501226]; Z.-z.~Xing, Phys.\ Lett.\ B {\bf
  618}, 141 (2005) [arXiv:hep-ph/0503200]; A.~Datta, L.~L.~Everett
and P.~Ramond, Phys.\ Lett.\ B {\bf 620}, 42 (2005)
[arXiv:hep-ph/0503222]; L.~L.~Everett, Phys.\ Rev.\ D {\bf 73}, 013011
  (2006) [arXiv:hep-ph/0510256].

\bibitem{Chauhan:2006im}
B.~C.~Chauhan, M.~Picariello, J.~Pulido and E.~Torrente-Lujan, Eur.\
Phys.\ J.\ C {\bf 50}, 573 (2007) [arXiv:hep-ph/0605032].

\bibitem{qlcRG}
A.~Dighe, S.~Goswami and P.~Roy, Phys. Rev. D {\bf 73},
  071301 (2006) [arXiv:hep-ph/0602062]; M.~A.~Schmidt and
  A.~Y.~Smirnov [arXiv:hep-ph/0607232].

\bibitem{qlcmodels}
T.~Ohlsson and G.~Seidl, Nucl.\ Phys.\ B {\bf 643}, 247 (2002) [arXiv:hep-ph/0206087]; P.~H.~Frampton and R.~N.~Mohapatra, JHEP {\bf 01}, 025
  (2005) [arXiv:hep-ph/0407139]; S.~Antusch, S.~F.~King and
  R.~N.~Mohapatra, Phys. Lett. B {\bf 618}, 150 (2005)
  [arXiv:hep-ph/0504007]; M.~Picariello, {\tt hep-ph/0611189};  A.~Hernandez-Galeana,
  Phys.\ Rev.\  D {\bf 76}, 093006 (2007)
  [arXiv:0710.2834 [hep-ph]].

\bibitem{Plentinger:2006nb}
  F.~Plentinger, G.~Seidl and W.~Winter,
  Nucl.\ Phys.\  B {\bf 791}, 60 (2008)
  [arXiv:hep-ph/0612169].

\bibitem{Plentinger:2007px}
  F.~Plentinger, G.~Seidl and W.~Winter,
  Phys.\ Rev.\  D {\bf 76}, 113003 (2007)
  [arXiv:0707.2379 [hep-ph]].

\bibitem{Winter:2007yi}
  W.~Winter,
  Phys.\ Lett.\  B {\bf 659}, 275 (2008)
  [arXiv:0709.2163 [hep-ph]].

\bibitem{Branco:2007nn}
  G.~C.~Branco, D.~Emmanuel-Costa, R.~Gonzalez Felipe and H.~Serodio,
  arXiv:0711.1613 [hep-ph];
  G.~C.~Branco, D.~Emmanuel-Costa, M.~N.~Rebelo and P.~Roy,
  arXiv:0712.0774 [hep-ph].

\bibitem{Froggatt:1978nt}
  C.~D.~Froggatt and H.~B.~Nielsen,
  Nucl.\ Phys.\  B {\bf 147}, 277 (1979).

\bibitem{discreteanomalies}
L.~M.~Krauss and F.~Wilczek, Phy.\ Rev.\ Lett.\ {\bf 62}, 1221 (1989).

\bibitem{Dreiner:2003yr}
 H.~K.~Dreiner, H.~Murayama and M.~Thormeier,
 Nucl.\ Phys.\  B {\bf 729} (2005) 278
 [arXiv:hep-ph/0312012]; H.~K.~Dreiner, C.~Luhn, H.~Murayama and M.~Thormeier,
 Nucl.\ Phys.\  B {\bf 774} (2007) 127
 [arXiv:hep-ph/0610026]; arXiv:0708.0989 [hep-ph].

\bibitem{Sayre:2007ps}
  J.~Sayre and S.~Wiesenfeldt,
  arXiv:0711.1687 [hep-ph].

\bibitem{Feruglio:2007hi}
  F.~Feruglio and Y.~Lin,
  arXiv:0712.1528 [hep-ph].

\bibitem{Antusch:2004yx}
  S.~Antusch, P.~Huber, J.~Kersten, T.~Schwetz and W.~Winter,
  Phys.\ Rev.\  D {\bf 70}, 097302 (2004)
  [arXiv:hep-ph/0404268].

\bibitem{Gandhi:2006gu}
  R.~Gandhi and W.~Winter,
  Phys.\ Rev.\  D {\bf 75}, 053002 (2007)
  [arXiv:hep-ph/0612158].

\bibitem{nonAbelian}
  F.~Plentinger and G.~Seidl, arXiv:0803.2889 [hep-ph].

\end{thebibliography}

\end{document}